\documentclass[showpacs,preprintnumbers,amssymb]{revtex4-2}
\usepackage{graphicx}
\usepackage{color}
\usepackage{bm}
\usepackage{amsmath}
\usepackage{amssymb}
\usepackage{epsfig}
\usepackage{amsfonts}
\usepackage{lineno,hyperref}
\usepackage{array}
\usepackage{float}
\usepackage{microtype}
\usepackage{multirow}
\usepackage{adjustbox}
\usepackage[english]{babel}
\usepackage{subfigure}
\usepackage{blindtext}
\usepackage[a4paper, total={7.5in, 10in}]{geometry}

\def \a{\alpha}
\def \b{\beta}
\def \l{\lambda}
\def \L{\Lambda}

\def \d{\delta}
\def \k{\kappa}

\def \be{\begin{equation}}
\def \ee{\end{equation}}
\def \ben{\begin{eqnarray}}
\def \een{\end{eqnarray}}
\def \o{\omega}

\def \p{\partial}

\def \G{\bar{G}}
\def \r{\rho}
\def \k{\kappa}
\def \R{\bar{R}}
\def \T{\bar{T}}

\def \T{\bar{T}}
\def \p{\bar{p}}
\def \La{\mathcal{L}}

\begin{document}
\title{Cosmological effects on $f(\bar{R},\bar{T})$ gravity through a non-standard theory}

\author{Arijit Panda}
\altaffiliation{arijitpanda260195@gmail.com}
\affiliation{Department of Physics, Raiganj University, Raiganj, Uttar Dinajpur-733 134, West Bengal, India. $\&$\\
Department of Physics, Prabhat Kumar College, Contai, Purba Medinipur-721404, India}

\author{Saibal Ray}
\altaffiliation{saibal.ray@gla.ac.in}
\affiliation{Centre for Cosmology, Astrophysics and Space Science (CCASS), GLA University, Mathura 281406, Uttar Pradesh, India.}

\author{Goutam Manna$^{a}$ }
\altaffiliation{goutamphs@pkcollegecontai.ac.in\\
$^{a}$Corresponding author}
\affiliation{Department of Physics, Prabhat Kumar College, Contai, Purba Medinipur 721404, West Bengal, India $\&$\\  Institute of Astronomy, Space and Earth Science (IASES), Kolkata 700054, West Bengal, India}

\author{Surajit Das}
\altaffiliation{surajit.cbpbu20@gmail.com}
\affiliation{Department of Physics, Cooch Behar Panchanan Barma University,
Panchanan Nagar, Vivekananda Street, Cooch Behar, West Bengal, India 736101}

\author{Chayan Ranjit}
\altaffiliation{chayanranjit@gmail.com}
\affiliation{Department of Mathematics, Egra S S B College, Egra, Purba Medinipur-721429, West Bengal, India}

\date{Received: date / Accepted: date}

\begin{abstract}
This study aims to investigate the impact of dark energy in cosmological scenarios by exploiting $f(\bar{R},\bar{T})$ gravity within the framework of a {\it non-standard} theory, called {\it {\bf K-}essence} theory, where $\bar{R}$ represents the Ricci scalar and $\bar{T}$ denotes the trace of the energy-momentum tensor associated with the {\bf K-}essence geometry. The Dirac-Born-Infeld (DBI) non-standard Lagrangian has been employed to generate the emergent gravity metric $(\bar{G}_{\mu\nu})$ associated with the {\bf K-}essence. This metric is distinct from the usual gravitational metric $(g_{\mu\nu})$. It has been shown that under a flat FLRW background gravitational metric, the modified field equations and the Friedmann equations of the $f(\bar{R},\bar{T})$ gravity are distinct from the usual ones. In order to get the equation of state (EOS) parameter $\omega$, we have solved the Friedmann equations by taking into account the function $f(\bar{R},\bar{T})\equiv f(\bar{R})+\lambda \bar{T}$, where $\lambda$ represents a parameter within the model. We have found a relationship between $\omega$ and time for different kinds of $f(\bar{R})$ by treating the kinetic energy of the {\bf K-}essence scalar field ($\dot{\phi}^{2}$) as the dark energy density which fluctuates with time. Surprisingly, this result meets the condition of the restriction on $\dot{\phi}^{2}$. By presenting graphical representations of the EOS parameter with time, we show that our model is consistent with the data of $SNIa$+$BAO$+$H(z)$ within a certain temporal interval.
\end{abstract}

\keywords{{\bf K-}essence emergent gravity, Modified theories of gravity, Friedman equations, Dark energy}
\pacs{04.20.-q, 04.50.-h, 04.50.Kd}        
\maketitle


\section{Introduction}\label{sec1}

The acceleration of the late-time universe is strongly evidenced by the discovery of Type Ia Supernovae (SNe Ia), Baryon Acoustic Oscillations (BAO), Cosmic Microwave Background (WMAP7), and the Planck mission \cite{Riess,Perlmutter,Komatsu,planck0,planck1,planck2}. In order to elucidate the phenomena known as the pushing-up effect, there has been speculation on the existence of an exotic entity referred to as {\it dark energy.} This speculation is based on an ad-hoc approach aimed at preserving the validity of established gravitational theories, particularly Einstein's general relativity (GR), by the inclusion of a cosmological constant. 

However, an alternative group of scientists has conducted a study on several aspects of gravitational theories and has proposed an explanation for the observed cosmic acceleration that does not rely on the existence of dark energy. Consequently, several theoretical ideas have been subjected to thorough scrutiny. In these models, the conventional Einstein-Hilbert action has been replaced with an arbitrary function of the Ricci scalar $R$. In their study, Carrol et al.  \cite{Carroll}  demonstrated the potential of $f(R)$ gravity to explain the cosmic acceleration noted in the latter stages of the universe.

Several pertinent theories, which include $f(R)$, $f(G)$ and $f(R,G)$, have been proposed in the literature \cite{Oikonomou3,Oikonomou9,Sotiriou,Felice,Nojiri1,Nojiri2,Nojiri3,Capozziello,Nojiri4,Elizalde1,Elizalde2,Myrzakulov,Durrer,Copeland}, which modify the traditional theory of gravity. In these theories, $G$ represents the Gauss-Bonnet invariant term. The criteria for accepting viable cosmological models have been examined in Capozziello's work \cite{Capozziello}.  The classical tests of GR in the Solar system regime have put strict constraints on the weak field, resulting in the elimination of a significant number of models that have been proposed so far \cite{Chiba,Olmo}. Nevertheless, some models have been documented in the literature \cite{Hu,Nojiri3} that have been deemed appropriate and have successfully undergone testing inside the Solar system. The authors of the following refs. \cite{Nojiri4,Nojiri5,Cognola,Odintsov3,Odintsov4,Odintsov5,Oikonomou6} have taken into account the $f(R)$ model, which provides a unified explanation for both inflation and the accelerating phase of the universe. This model is also consistent with local tests conducted inside the Solar system. The possibility of comprehending the galactic dynamics of large test particles without invoking dark matter was demonstrated in some previous studies \cite{Capozziello2,Borowiec} within the framework of $f(R)$ gravity models. Interested readers are encouraged to refer to the following articles \cite{Nojiri6,Nojiri9,Sotiriou} which offer comprehensive evaluations of $f(R)$ generalized gravity models. 

A more specific application, known as $f(R, \La_m)$ gravity, was put forward in \cite{Berto,Harko1} which is dependent on the concept of least action. It may be seen as a relativistically invariant model of interacting dark energy. A description of $f(R)$ models in the context of {\bf K-}essence geometry, the $f(\bar{R},\La(X))$ gravity has recently been developed by Manna et al. \cite{Manna} based on the Dirac-Born-Infeld (DBI) variety of action. Here $\bar{R}$ is the Ricci scalar and $\La(X)$ is the DBI type non-canonical Lagrangian with the expression $X=\frac{1}{2}g^{\mu\nu}\nabla_{\mu}\phi\nabla_{\nu}\phi$ with $\phi$ being the scalar field of the {\bf K-}essence geometry. The cosmological constant present in the gravitational Lagrangian is nothing but a function of the trace of the stress-energy tensor. It is to note that more often this model is denoted as ``$\Lambda (T)$ Gravity" \cite{Poplawski,Sahoo3}. Without defining the precise shape of the function $\Lambda (T)$, it was achieved that the existing cosmological data reflect on a changing cosmological constant that is compatible with $\Lambda (T)$ gravity  \cite{Poplawski}.

The concept of $f(R,T)$ gravity was initially introduced by Harko et al. \cite{Harko}. The observed phenomenon of extending $f(R)$ gravity, where $T$ represents the trace of the stress-energy tensor, is also evident in this particular scenario. When considering the stress-energy tensor as a source, the trajectory of particles deviates from the geodesic path due to the presence of an extra force acting perpendicular to the four-velocity of the test particle. The precession of Mercury's perihelion has been utilized to establish a general constraint on the size of any new acceleration. Harko et al. \cite{Harko} employed the Hilbert-Einstein type variational technique to compute the covariance divergence of the stress-energy tensor and the field equations within their model. The model is dependent on a source term, which corresponds to the matter Lagrangian ($\La_{m}$). Hence, it may be inferred that distinct Lagrangians will provide diverse field equations. They \cite{Harko} have also examined other prominent models pertaining to various $f(R)$ alternatives, such as the scalar field model known as $f(R, T)$.

The cosmic acceleration is connected to both the composition of matter and the influence of geometrical parameters, as stated by Harko et al. \cite{Harko}.  Upon rectifying the conservation rule proposed in the work \cite{Harko}, Barrientos et al. \cite{Barrientos} observed a discrepancy in the equation of motion for a test particle. This discrepancy indicates that the motion of the particle deviates from geodesic motion, even in the absence of pressure. According to the principle of conservation of the stress-energy tensor, it has been postulated in the work of Alvarenga et al. \cite{Alvarenga} that the stress-energy tensor, denoted as $T$, possesses a specific configuration and is not subject to arbitrary selection. The thermodynamics of this model have been extensively examined in the study conducted by Sharif et al. \cite{Sharif2}. 
 
Bianchi-III and Bianchi$-VI_0$ cosmological models with string fluid source in $f(R,T)$ gravity have been explored by Sahoo et al. \cite{Sahoo}.  They have examined some of the dynamical and physical behaviors of their model and constructed the field equations utilizing a time-varying deceleration parameter. The first and second classes of $f(R,T)$ gravity applicable to the anisotropic Bianchi type-I space-time have been studied by Singh and Bishi \cite{Singh2,Singh3}. In order to solve the field equations, they have also addressed the polynomial and exponential power law expansions as well as the generalized scale factor. Continuing their work, Harko et al. \cite{Harko} and other scientists looked into the $f(R,T)$ concept for various matter distributions \cite{Baffou,Aygun,Sahu,Satish}. The dynamical system of the $f(R,T)$ gravity has been explored in this context by Mirza et al. \cite{Mirza}. Presuming energy conservation, equations of motion and possible future singularities for a barotropic perfect fluid and a fluid resembling dark energy have been considered. In this study, the authors also discovered that there is no future singularity for the barotropic fluid, although other types of singularities, such as the fluid may arise for dark energy owing to the additional degrees of freedom in specific options of the equation of state. Different cosmological scenarios with $f(R,T)$ gravity are discussed in the literatures \cite{Vainshtein,Horndeski,Fradkin,Vasiliev,Khoury,Khoury1,Nojiri8,Deffayet,Sharif,Houndjo,Houndjo2,Jamil,Alvaro,Clifton,Houndjo3,Zubair,Singh,Zubair1,Zubair2,Lobo3,Moraes,Hulke2020,Zubair3,Rosa,Singh4,Rosa2}.

The authors Manna et al. \cite{gm1,gm2,gm3}  have constructed an essential form of the emergent gravitational metric $\bar{G}_{\mu\nu}$ using the DBI action \cite{Mukohyama,Born,born2,Dirac}. This version of the metric is distinct from the conventional gravitational metric $g_{\mu\nu}$ and lacks manifest conformal equivalence. It should be acknowledged that the {\bf K-}essence theory is a non-traditional scalar field theory.  According to the idea of {\bf K-}essence, the dominance of kinetic energy over potential energy is observed in the scalar field. This observation has been supported by several studies \cite{Visser,Babichev2,Vikman,Babichev4,Chimento}. It is important to acknowledge that within this particular setting, the {\bf K}-essence model serves as one of the models employed for investigating the effects of dark energy in the present-day cosmos. The dynamical solutions of the {\bf K}-essence equation of motion, which exhibit both spontaneous Lorentz invariance breaking and metric alterations for the corresponding perturbations, serve as the distinguishing factor between the {\bf K}-essence theory featuring non-canonical kinetic terms and relativistic field theories with canonical kinetic terms. The theoretical expression of the {\bf K}-essence field Lagrangian exhibits non-canonical characteristics, resulting in perturbations that propagate throughout the {\it emergent} curved spacetime, sometimes referred to as the {\it analogue} spacetime, as required by the metric. The Lagrangian of {\bf K-}essence model has a form $\La =-V (\phi)F(X)$ where $\phi$ is the {\bf K-}essence scalar field and $X =\frac{1}{2} g_{\mu\nu} \nabla^{\mu} \phi \nabla^{\nu} \phi$.  However, there is a different Lagrangian form \cite{Tian} where the Lagrangian is employed as arbitrary functions of $\phi$ and $X$. Based on the findings of the Planck collaboration, namely the $2015-XIV$ \cite{planck0} and $2018-VI$ \cite{planck2}, it may be concluded that this theory possesses significant empirical evidence. 

Let us also look at the non-canonical Lagrangian from a different angle which is as follows: \\ In general, we may define Lagrangian in canonical or standard form as $\La=T-V$, where $T$ is the kinetic energy and $V$ is the potential energy of the system. However, as stated by Goldstein and Rana \cite{Goldstein,Rana}, the general form of the Lagrangian is non-canonical, whereas the canonical form is obtained under certain conditions. In scleronomic systems, the forces cannot be deduced from a potential function and hence indicate that the canonical Lagrangian cannot exhibit any explicit dependency on time. Again, for the systems subject to dissipative processes all scleronomic systems are not necessarily conservative. We can readily derive the canonical Lagrangian from the non-canonical one. The uniqueness of the functional form of $\La$ is fundamentally not guaranteed, as the Euler-Lagrange equations of motion can be retained for various Lagrangian choices \cite{Goldstein,Rana}. Moreover, Raychaudhuri \cite{Raychaudhuri} highlights that when extending our analysis outside the realm of mechanics, the conventional notions of kinetic and potential energies become inadequate. Consequently, the expression $\La=T-V$ ceases to hold true. The initial impetus for this endeavor likely originated from a process of retroactive computation, wherein the field equations were already established, needing the determination of an appropriate Lagrangian density to ensure their accurate representation. Furthermore, it is worth noting that the conventional understanding of the classical idea denoted as $\La (=T-V)$ is no longer valid within the framework of special relativistic dynamics. Consequently, it may be asserted that the general form of the Lagrangian is of a non-canonical nature \cite{Das}.

Now we would like to discuss the importance of the {\bf K-}essence theory with a focus on the Lagrangian of the DBI type: the observational findings of large-scale structure, searches for type Ia supernovae, and studies of cosmic microwave background anisotropy \cite{Bahcall} provide strong evidence for the acceleration of the cosmos. It is very interesting that the majority of scientists also think that dark energy, which exerts a negative pressure, predominates in our universe. A possibility of such an exotic component of the universe was previously suggested by scientists as the Einsteinian cosmological constant \cite{Einstein1918} or the Zel`dovich vacuum density \cite{Zeldovich1967,Zeldovich1968}. 
It is important to acknowledge that the canonical model, such as the $\Lambda$CDM model, is unable to provide a satisfactory explanation for both the fine-tuning and coincidence issues. This phenomenon can be elucidated as described by Yoo \cite{Yoo}. At the Planck scale, denoted by $\hbar=1$, the Planck mass is estimated to be around $M_{pl}\sim (8\pi G)^{-1/2}\sim 10^{18}GeV$. Consequently, the energy density $\rho_{\L}$ may be approximated as $(10^{18}GeV)^{4}\sim 2\times 10^{110} erg/cm^{3}$. Nevertheless, the majority of cosmological observations suggest that the observed energy density of the cosmological constant, denoted as $\rho_{\Lambda}^{Obs}$, is less than or equal to $(10^{-12}GeV)^{4}$, which is approximately equivalent to $2\times 10^{-10} erg/cm^{3}$. There exists a disparity of 120 orders of magnitude between the anticipated theoretical outcome and the observed value \cite{Adler1995,Bengochea2020}. 
The aforementioned disparity has been referred to as "the most awful theoretical prediction in the history of physics."
 The inquiry pertains to the underlying reasons for the small magnitude of vacuum energy. Can it effectively nullify a factor of $10^{120}$? The following topics represent significant unresolved matters within the fields of physics and cosmology. Another issue pertaining to the cosmological constant is the inquiry into why the energy density associated with it, denoted as $\rho_{\L}$, is not only significantly small but also comparable in magnitude to the current mass density of the Universe. In alternative terms, what justifies the commencement of cosmic acceleration at the present moment as opposed to a future time?  Nevertheless, inside the K-essence model, the behavior of the K-essence field is solely influenced by the radiation background. As a result, it avoids the need for fine-tuning that was present in the canonical model. Furthermore, the coincidence problem is also resolved by the presence of an S-attractor that attracts shortly after the beginning of the matter-dominated phase.
Additionally, observable findings suggest that the universe is mostly flat in space. The current universe seems to contain roughly $70\%$ dark energy (DE), which is one of the causes of cosmic acceleration. We utilize the equation of state (EoS) parameter $\omega$, which is described by the equation $p=\omega \rho$, to characterize the behavior of the matter-energy density of any given matter-energy component. For non-relativistic matter, the EoS parameter should be set to $0$, for radiation to be set to $\frac{1}{3}$ and for dark energy dominated epoch to be set to $-1$ \cite{wienberg}. But in principle, it can have any value, and it can change with respect to time. 

The {\bf K-}essence theory \cite{Visser,Babichev2,Babichev4,Vikman,Picon1,Bahcall,Picon2,Scherrer1,scherrer2,Picon3,Chimento}, a new class of scalar field models with intriguing dynamic features, was introduced at this point which successfully could solve the fine-tuning issue. The most crucial aspect of this theory is that the scalar field's nonlinear kinetic energy is the sole source of negative pressure. There are several theories that include attractor solutions \cite{Picon2,Kang}, in which the scalar field modifies the rate of evolution to create the {\bf K-}essence theory's equation of state at various epochs in accordance with the background's changing equation of state. The ratio of {\bf K-}essence field to the radiation density remained constant throughout the radiation-dominated period because of the fact that the {\bf K-}essence field sub-dominated and duplicated the radiation's equation of state (EoS). Due to dynamical constraints, the {\bf K-}essence field was unable to replicate the dust-like EoS at the time of the dust-dominated epoch, but it rapidly dropped down from its energy value by many orders of magnitude and acquired a constant value. Later, at a point approximately equivalent to the present age of the universe, the {\bf K-}essence field suppressed the matter density and consequently, the cosmos entered the cosmic acceleration period. Finally, the EoS of the {\bf K-}essence theory gradually reverts to a value between $0$ and $-1$.

Another intriguing aspect of the {\bf K-}essence concept is its ability to generate a dark energy component in which the sound speed ($c_{s}$) is always slower than the light speed. This property may lessen the cosmic microwave background (CMB) disturbances on large angular scales \cite{eric,dedeo,bean}. These models are observationally distinct from the usual scalar field quintessence models with a classical kinetic component (for which $c_{s} = 1$). However, there are several phases when the fluctuations of the field may spread superluminally ($c_{s}>1$) \cite{Babichev2,Vikman,Bonvin}. A few cosmological behaviors and the stability of the {\bf K-}essence model in FLRW space-time have been examined by Yang et al. \cite{Yang}. Results that are out of the ordinary for modest sound speed of scalar perturbations show that dark energy is clustering and cosmic perturbations are increasing \cite{Sawicki,Kunz}.

In \cite{Bandyopadhyay,Planck1}, one may find the observable data supporting the {\bf K-}essence theory whether paired with a scalar field, a modified gravity theory, or both. The scale factor function, which relates to a non-minimally linked {\bf K-}essence model, was confined by the observational data (a detailed discussion and analysis can be found in the ref. \cite{Planck1}). 
In order to eliminate the infinite self-energy of the electron, Born and Infeld \cite{born2} historically suggested a theory with a non-canonical kinetic term. The available literature \cite{Dirac} also examined hypotheses of a similar sort. The DBI type non-canonical Lagrangian has also been employed by several scientists \cite{Das,Linde,Albrecht,Dvali,Kachru,Alishahiha,Silverstein1,Chen1,Weinberg,Chen2}, specific examples of which are string theory, brane cosmology, D-branes, and other related topics.

In the present study, we have used the {\bf K-}essence model to examine the $f(\R,\T)$ theory of gravity. A more extended $f(\R,\T)$ theory was evaluated by the metric formalism in the {\bf K-}essence geometry, where $\R$ is the Ricci scalar and $\T$ is the trace of stress-energy tensor of this geometry based on the DBI type non-canonical Lagrangian $\La(X)$ \cite{Visser,Babichev2,Babichev4,Vikman,gm1,gm2,gm3,gm4}. For various selections of $f(\R,\T)$, we have obtained the modified Friedmann equations in the {\bf K-}essence geometry. From these equations, we have obtained the associated pressure and energy density for each form of $f(\R,\T)$. Then, for each of these several models, we have estimated the pressure ($\p$) and energy density ($\bar{\r}$). We have displayed the fluctuation of the EoS parameter over time while keeping in mind its definition. Additionally, we have compared the estimates from those graphs to the recent observational data of $SNIa+BAO+H(z)$ \cite{Tripathi}.

The utilization of the non-standard technique in the investigation of gravitation and cosmology is primarily motivated by the reasons outlined afterward. The conventional, canonical, or standard theories do not provide a comprehensive understanding of the subject matter. The current understanding of cosmology
lacks precise explanations for phenomena such as dark matter, dark energy, the Big Bang, the matter-antimatter inequalities, the cosmological constant problem, the size and form of the universe, cosmic inflation, the horizon problem, and other related aspects. The primary unresolved issue in the field
of basic physics is the reconciliation of gravity and quantum mechanics within
a unified theoretical framework. There remain unfinished tasks that require
attention. In light of the aforementioned reasons, we have chosen to employ the non-standard or non-canonical approach for our study, specifically utilizing the {\bf K-}essence theory. Also, we know that Einstein's equation has two parts, the LHS says about the geometry and the RHS says about the matter-energy. Actually, Einstein's work had been constructed on a simple and specific basis. However, physicists always intend to construct a generalized theory in any field of study. Over the past years, there has been modification mainly in either part of the equation. $f(R)$ theories are brought into the light to modify the curvature term ($R$) of Einstein's equation and different types of Lagrangian try to generalize the matter part of the equation. But it is evident from the works of \cite{Goldstein,Rana,Raychaudhuri} that the most general Lagrangian should be non-canonical. It is worth mentioning that the frequently used modified gravity theories, such as $f(R)$ or $f(R, T)$ gravity, as well as the K-essence theory, are utilized both in the context of the usual gravitational phenomena and in the investigation of dark energy and dark matter. Our study relies on the investigation of the potential consequences that may arise from concurrently modifying both the curvature component and the matter component of Einstein's equation, by including modified gravity ($f(R,T)$) and non-canonical Lagrangian.  In this context, here, we investigate the $f(\R, \T)$ gravity through the {\bf K-}essence geometry.

This paper is embodied in the following way:  In Section II, we have briefly discussed about the {\bf K-}essence emergent geometry with the help of the information available in the following literature \cite{gm1,gm2,gm3,Visser,Babichev2,Vikman}. In Section III, we have developed the modified field equations and corresponding Friedmann equations of $f(\R,\T)$-gravity in the {\bf K-}essence geometry considering the background gravitational metric to be flat FLRW type. We have discussed the variations of EOS parameters for different choices of $f(\R,\T)$ in Section IV. Also, we have matched our results with the observational data from \cite{Tripathi}. The last Section V is devoted to the conclusion and discussion of our work. Additionally, we have provided the full expression of the EOS parameters in {\bf Appendix A}.

\section{Brief review of {\bf K-}essence theory}
To give an overview of the {\bf K-}essence geometry, let us start by the following action  \cite{Babichev2,Babichev4}:
\ben
S_{k}[\phi,g_{\mu\nu}]= \int d^{4}x {\sqrt -g} \La(X,\phi),
\label{1}
\een
where $X=\frac{1}{2}g^{\mu\nu}\nabla_{\mu}\phi\nabla_{\nu}\phi$ is the canonical kinetic term and $\La(X,\phi)$ is the non-canonical Lagrangian. Here, the {\bf K-}essence scalar field $\phi$ has coupled minimally with the usual gravitational metric $g_{\mu\nu}$. 

The energy-momentum tensor is defined as:
\ben
T_{\mu\nu}\equiv \frac{-2}{\sqrt {-g}}\frac{\delta S_{k}}{\delta g^{\mu\nu}}=-2\frac{\partial \La}{\partial g^{\mu\nu}}+g_{\mu\nu}\La=-\La_{X}\nabla_{\mu}\phi\nabla_{\nu}\phi
+g_{\mu\nu}\La, 
\label{2}
\een
where $\La_{\mathrm X}= \frac{d\La}{dX},~\La_{\mathrm XX}= \frac{d^{2}\La}{dX^{2}},
~\La_{\mathrm\phi}=\frac{d\La}{d\phi}$ and $\nabla_{\mu}$ is the covariant derivative defined with respect to the gravitational metric $g_{\mu\nu}$. 

The scalar field equation of motion (EOM) is
\ben
-\frac{1}{\sqrt {-g}}\frac{\delta S_{k}}{\delta \phi}= \tilde{G}^{\mu\nu}\nabla_{\mu}\nabla_{\nu}\phi +2X\La_{X\phi}-\La_{\phi}=0,
\label{3}
\een
where  
\ben
\tilde{G}^{\mu\nu}\equiv \frac{c_{s}}{\La_{X}^{2}}[\La_{X} g^{\mu\nu} + \La_{XX} \nabla ^{\mu}\phi\nabla^{\nu}\phi]
\label{4}
\een
with $1+ \frac{2X\La_{XX}}{\La_{X}} > 0$ and $c_s^{2}(X,\phi)\equiv{(1+2X\frac{\La_{XX}} {\La_{X}})^{-1}}$.

The inverse metric is
\ben G_{\mu\nu}=\frac{\La_{X}}{c_{s}}[g_{\mu\nu}-{c_{s}^{2}}\frac{\La_{XX}}{L_{X}}\nabla_{\mu}\phi\nabla_{\nu}\phi].
\label{5}
\een

After a conformal transformation \cite{gm1,gm2} $\bar G_{\mu\nu}\equiv \frac{c_{s}}{\La_{X}}G_{\mu\nu}$ we have
\ben
\bar{G}_{\mu\nu}=g_{\mu\nu}-\frac{\La_{XX}}{\La_{X}+2X\La_{XX}}\nabla_{\mu}\phi\nabla_{\nu}\phi.
\label{6}
\een

The Eqs. (\ref{4})--(\ref{6}) are physically relevant if $\La_{X}\neq 0$ for a positive definite $c_{s}^{2}$. Basically, Eq. (\ref{6}) states that our emergent metric $\bar{G}_{\mu\nu}$ is conformally distinct from $g _{\mu\nu}$ for non-trivial spacetime configurations of $\phi$. Like canonical scalar fields, $\phi$ has varied local causal structural features. It is also distinct from those defined with $g_{\mu\nu}$. The equation of motion Eq. (\ref{3}) becomes relevant if the non-explicit reliance of $\La$ on $\phi$ can be addressed. Then the EOM Eq. (\ref{3}) is:
\ben
\frac{1}{\sqrt{-g}}\frac{\delta S_{k}}{\delta \phi}= \bar G^{\mu\nu}\nabla_{\mu}\nabla_{\nu}\phi=0.
\label{7}
\een

Considering the DBI type non-canonical Lagrangian $\La(X,\phi)\equiv \La(X)$  \cite{gm1,gm2,gm3,Mukohyama,Born,born2,Dirac}:
\ben
\La(X)= 1-\sqrt{1-2X}.
\label{8}
\een

The {\bf K-}essence geometry states that kinetic energy dominates over the potential energy and hence the potential term in our Lagrangian Eq. (\ref{8}) \cite{Mukohyama} is cut out and $c_{s}^{2}$ is $(1-2X)$. Thus, the effective emergent metric, i.e. Eq. (\ref{6}), becomes
\ben
\bar G_{\mu\nu}= g_{\mu\nu} - \nabla_{\mu}\phi\nabla_{\nu}\phi= g_{\mu\nu} - \partial_{\mu}\phi\partial_{\nu}\phi,
\label{9}
\een
since $\phi$ is a scalar. 

Following \cite{gm1,gm2}, the Christoffel symbol associated with the emergent gravity metric Eq. (\ref{9}) is: 
\ben
\bar\Gamma ^{\alpha}_{\mu\nu} 
&&=\Gamma ^{\alpha}_{\mu\nu} -\frac {1}{2(1-2X)}\Big[\delta^{\alpha}_{\mu}\partial_{\nu}
+ \delta^{\alpha}_{\nu}\partial_{\mu}\Big]X,~~~~~~~~~~~
\label{10}
\een
where $\Gamma ^{\alpha}_{\mu\nu}$ is the usual Christoffel symbol associated with the gravitational metric $g_{\mu\nu}$.

Therefore, the geodesic equation for the {\bf K-}essence geometry becomes:
\ben
\frac {d^{2}x^{\alpha}}{d\l^{2}} +  \bar\Gamma ^{\alpha}_{\mu\nu}\frac {dx^{\mu}}{d\l}\frac {dx^{\nu}}{d\l}=0, \label{11}
\een
where $\l$ is an affine parameter.

The covariant derivative $D_{\mu}$ \cite{Babichev2} linked with the emergent metric $\bar{G}_ {\mu\nu}$ $(D_{\a}\bar{G}^{\a\b}=0)$ yields
\ben
D_{\mu}A_{\nu}=\partial_{\mu} A_{\nu}-\bar \Gamma^{\l}_{\mu\nu}A_{\l}, \label{12}
\een
and the inverse emergent metric is $\bar G^{\mu\nu}$ such as $\bar G_{\mu\l}\bar G^{\l\nu}=\delta^{\nu}_{\mu}$.

Therefore, if we consider the total action which describes the dynamics of {\bf K}-essence and General Relativity \cite{Vikman}, then the ``Emergent Einstein's Field Equation (EEFE)'' reads:
\ben
\bar{\mathcal{G}}_{\mu\nu}=\R_{\mu\nu}-\frac{1}{2}\bar{G}_{\mu\nu}\R=\k \T_{\mu\nu}, \label{13}
\een
where $\k=8\pi G$ is constant, $\R_{\mu\nu}$ is the Ricci tensor and $\R~ (=\R_{\mu\nu}\bar{G}^{\mu\nu})$ is the Ricci scalar and $\T_{\mu\nu}$ is the energy-momentum tensor of the emergent spacetime. It is important to acknowledge that the aforementioned EEFE is only derived from the framework of {\bf K}-essence geometry. If the {\bf K}-essence scalar field ($\phi$) is removed from this geometry, the resulting equation becomes equivalent to the conventional Einstein field equation. In the present setting, it is plausible to assert that both our understanding of geometry and the Einstein field equation are subject to modification.

\section{\texorpdfstring{$f(\R,\T)$}{TEXT} Gravity in {\bf K-}essence geometry}
In this section, we describe the $f(\R,\T)$ gravity in the context of {\bf K-}essence geometry in a similar manner to Harko et all. \cite{Harko}. For this, first, we take the action of the modified gravity in the context of the {\bf K-}essence geometry ($\k=1$) as
\ben
S=\int d^4x\sqrt{-\G}\Big[f(\R,\T)+\La(X)\Big],
\label{14}
\een
where $f(\R,\T)$ is an arbitrary function of the Ricci scalar ($\R$) and trace of the energy-momentum tensor $(\T=\T^{\mu\nu}\G_{\mu\nu})$ and $L(X)$ is the non-canonical Lagrangian corresponding to the {\bf K}-essence theory. Our revised action (\ref{14}) is clearly dependent on $\R$, $\T$, and $X(=\frac{1}{2}g^{\mu\nu}\nabla_{\mu}\phi\nabla_{\nu}\phi)$, rather than on the {\bf K-}essence scalar field ($\phi$) explicitly. We can define the emergent energy-momentum tensor of this geometry as
\ben
\T_{\mu\nu}=-\frac{2}{\sqrt{-\G}}\frac{\partial\Big(\sqrt{-\G}\La(X)\Big)}{\partial \G_{\mu\nu}}
\label{15}
\een
where $\big(-\G \big)^{1/2}=\big(-det({\G_{\mu\nu}})\big)^{1/2}$.

It is important to note that the emergent energy-momentum tensor (\ref{15}) may be defined in light of the modifications to the geometry and the Einstein field equation (\ref{13}). It is also worth noting that the calculation of the emergent energy-momentum tensor ($\T_{\mu\nu}$) may be performed by two methods: the first one involves directly calculating the left-hand side of Einstein's equations for emergent gravity (\ref{13}), while the other entails utilizing the definition of the emergent energy-momentum tensor (\ref{15}). The objective of our present study is to develop the $f(\bar{R},\bar{T})$ gravity model within the framework of a non-canonical theory, specifically the {\bf K-}essence scalar field theory. The energy-momentum tensor $(\bar{T}_{\mu\nu})$ that arises in the  EEFE for the emergent gravity theory is specifically associated with the {\bf K-}essence theory. Additionally, the emergent energy-momentum tensor in Eq. (\ref{15}) comes from the formulation of the action principle in the $f(\bar{R},\bar{T})$ gravity model, constructed in the background of {\bf K-}essence emergent geometry. So, the effect of the emergent energy-momentum tensor of $f(\bar{R},\bar{T})$ gravity model in the context of {\bf K-}essence theory can be observed. For this, we use the definition of emergent energy-momentum tensor (\ref{15}) throughout the work. Furthermore, it is important to acknowledge that under the framework of modified gravity theory, as put out by Harko et al. \cite{Harko}, the stress-energy tensor is contingent just upon the matter Lagrangian. Additionally, it is assumed that the matter Lagrangian is reliant solely on the metric tensor, rather than its derivatives. So, the stress-energy tensor can not depend on the function $f(\bar{R},\bar{T})$ or more specifically on $\bar{R}$. Therefore, the analysis of the variation of the function $L(X)$ has been undertaken in this context.

Considering $\bar{G}_{\mu\nu}$ as in Eq. (\ref{9}), the emergent energy-momentum tensor (\ref{14}) can also be written as 
\ben
\T_{\mu\nu}
=\G_{\mu\nu}\La(X)-2\frac{\partial \La(X)}{\partial \G^{\mu\nu}}.
\label{16}
\een

Varying the action  we get
\ben 
\delta S=\int \Big[F(\R,\T)\delta\R+f_{\T}(\R,\T)\frac{\delta \T}{\delta\G^{\mu\nu}}\delta\G^{\mu\nu}+\delta\sqrt{-\G}+\frac{1}{\sqrt{-\G}}\frac{\partial(\sqrt{-\G}\La(X))}{\partial \G^{\mu\nu}}\Big]\sqrt{-\G}d^4x,\nonumber\\
\label{17}
\een
where $F(\R,\T)=\partial f(\R,\T)/\partial \R\equiv F$ and $f_{\T}(\R,\T)=\partial f(\R,\T)/\partial \T\equiv f_{\T}$ respectively. 

The expression of $\R_{\mu\nu}$ and $\bar{\Gamma}_{\mu\nu}$ in this geometry are given by
\ben
\bar{R}_{\mu\nu}=\bar{R}^{\a}_{\mu\a\nu}=\partial_{\a}\bar{\Gamma}^{\a}_{\mu\nu}-\partial_{\nu}\bar{\Gamma}^{\a}_{\mu\a}+\bar{\Gamma}^{\rho}_{\mu\nu}\bar{\Gamma}^{\a}_{\rho\a}-\bar{\Gamma}^{\rho}_{\mu\a}\bar{\Gamma}^{\a}_{\nu\rho},
\label{18}
\een

\ben
\bar{\Gamma}^{\alpha}_{\mu\nu} = \frac{1}{2}\bar{G}^{\alpha\beta}\Big[\partial_{\mu}\bar{G}_{\beta\nu} + \partial_{\nu}\bar{G}_{\mu\beta} - \partial_{\beta}\bar{G}_{\mu\nu}\Big].
\label{19}
\een

The variation of the Ricci scalar for the {\bf K}-essence emergent spacetime is \cite{Manna,Harko1,Harko,Sahoo3}:  
\ben
\d\R=\R_{\mu\nu}\d\G^{\mu\nu}+\G_{\mu\nu}D_{\a}D^{\a}\d\G^{\mu\nu}-D_{\mu}D_{\nu}\d\G^{\mu\nu}.
\label{20}
\een

Putting Eq. (\ref{20}) in Eq. (\ref{17}), we get 
\ben 
\delta S&=&\int \Bigg[F(\R,\T)\R_{\mu\nu}\delta\G^{\mu\nu}+F(\R,\T)\G_{\mu\nu}\bar{\square} \delta \G^{\mu\nu}
-f(\R,\T)D_{\mu}D_{\nu}\delta \G^{\mu\nu}\nonumber\\&&+f_{\T}(\R,\T)\frac{\delta \T}{\delta\G^{\mu\nu}}\delta\G^{\mu\nu}
-\frac{1}{2}\G^{\mu\nu}\partial \G^{\mu\nu}F(\R,\T)
+\frac{1}{\sqrt{-\G}}\frac{\partial(\sqrt{-\G}\La(X))}{\partial \G^{\mu\nu}}\Bigg]\sqrt{-\G}d^4x,\nonumber\\
\label{21}
\een
where $\bar{\square}=D_{\mu}D^{\mu}$ and $D_{\mu}$ is the covariant derivative with respect to the metric $\bar{G}_{\mu\nu}$.

Again following \cite{Harko}, we have
\ben
\frac{\partial \T}{\partial \G^{\mu\nu}}=\frac{\partial(\T_{\alpha\beta}\G^{\alpha\beta})}{\partial \G^{\mu\nu}}=\T_{\mu\nu}+\bar{\Theta}_{\mu\nu},
\label{22}
\een
where
\ben
\bar{\Theta}_{\mu\nu}=\G^{\alpha\beta}\frac{\partial \T_{\alpha\beta}}{\partial\G^{\mu\nu}}.
\label{23}
\een

Using Eqs. (\ref{15}), (\ref{21}), (\ref{23}) and applying the least action principle we obtain the modified field equation as 
\ben
&&F\R_{\mu\nu}-\frac{1}{2}f(\R,\T)\G_{\mu\nu}+(\G_{\mu\nu}\bar{\square}-D_{\mu}D_{\nu})F=\frac{1}{2}\T_{\mu\nu}-f_{\T}\T_{\mu\nu}-f_{\T}\bar{\Theta}_{\mu\nu}.~~~~~~\label{24}
\een
The modified field equation (\ref{24}) looks like the field equations of standard $f(R,T)$ gravity, as derived by Harko \cite{Harko} (Eq. (11)), Sahoo \cite{Sahoo3} (Eq. 4), and other authors. However, it differs significantly in terms of structure development when considering the viewpoint of {\bf K-}essence geometry. One can easily get back those equations if one ignores the {\bf K-}essence scalar field-related term and $\La(X)=\La_{m}$. Contracting Eq. (\ref{24}) with $\G^{\mu\nu}$ we get,
\ben
F\R-2f(\R,\T)+3\bar{\square}F=\frac{1}{2}\T-f_{\T}\T-f_{\T}\bar{\Theta}
\label{24.1}
\een
where $\bar{\Theta}=\bar{\Theta}_{\mu}^{~\mu}$.\\
We can easily cut off the term $\bar{\square}F$ using Eq.(\ref{24}) and Eq.(\ref{24.1}) as
\ben
&&F(\R_{\mu\nu}-\frac{1}{3}\R\G_{\mu\nu})+\frac{1}{6}f(\R,\T)=\frac{1}{2}(\T_{\mu\nu}-\frac{1}{3}\T\G_{\mu\nu})\nonumber\\
&&-f_{\T}(\T_{\mu\nu}-\frac{1}{3}\T\G_{\mu\nu})-f_{\T}(\bar{\Theta}_{\mu\nu}-\frac{1}{3}\bar{\Theta}\G_{\mu\nu})+D_{\mu}D_{\nu}F
\label{24.2}
\een
Consistent with the findings of Harko et al. \cite{Harko} and Koivisto \cite{Koivisto}, the covariant derivative of the field equation (\ref{24}) in this particular geometry is likewise determined to be zero. This implies that according to the geometrical identity established by Koivisto \cite{Koivisto} $D^{\mu}\G_{\mu\nu}=0$ and $(\bar{\square}D_{\nu}-D_{\nu}\bar{\square})F=0$, we obtain
\ben
&&D^{\mu}\Big[F\R_{\mu\nu}-\frac{1}{2}f(\R,\T)\G_{\mu\nu}+(\G_{\mu\nu}\bar{\square}-D_{\mu}D_{\nu})F
\Big]=0.
\label{24.3}
\een
Furthermore, we have the condition for the preservation of the energy-momentum tensor ($D^{\mu}\T_{\mu\nu}=0$) as \cite{Harko,Koivisto,Carvalho}
\ben
D^{\mu}\T_{\mu\nu}=\frac{f_{\T}}{\frac{1}{2}-f_{\T}}\Big[(\T_{\mu\nu}+\bar{\Theta}_{\mu\nu})D^{\mu}(\ln f_{\T})+D^{\mu}\bar{\Theta}_{\mu\nu}\Big].
\label{25}
\een

Consider the flat Friedmann-Lema{\i'}tre-Robertson-Walker (FLRW) metric as a background gravitational metric ($g_{\mu\nu}$) and the line element is
\ben
ds^{2}=dt^{2}-a^{2}(t)\sum_{i=1}^{3} (dx^{i})^{2}, \label{26}
\een
where $a(t)$ is the usual scale factor. 

Reminding Eq. (\ref{9}), we can write the components of the emergent gravity metric as
\ben
\G_{00}=(1-\dot\phi^{2})~;~\G_{ii}=-(a^{2}(t)+(\phi^{'})^{2})~;
~\G_{0i}=-\dot\phi \phi^{'}=\G_{i0}, \label{27}
\een
where we consider $\phi\equiv \phi(t,x^{i})$, $\dot\phi=\frac{\partial \phi}{\partial t}$ and $\phi^{'}=\frac{\partial \phi}{\partial x^{i}}$.

Admitting the homogeneity of {\bf K-}essence scalar field we choose it to be a function of time only, i.e., $\phi=\phi(t)$ \cite{Manna,gm4,gm6,gm7} so that $\partial_{\rho}\phi\partial^{\rho}\phi=D_{\rho}\phi D^{\rho}\phi=\dot{\phi}^2$. As the dynamical solutions of the {\bf K}-essence scalar fields spontaneously break the Lorentz symmetry, it is meaningful to take the choice as mentioned above. 

Therefore, following \cite{Manna,gm6}, we can write the {\bf K-}essence emergent line element as 
\ben
dS^{2}=(1-\dot\phi^{2})dt^{2}-a^{2}(t)\sum_{i=1}^{3} (dx^{i})^{2},
\label{28}
\een
and from the EOM Eq. (\ref{7}), we can get the relation between the Hubble parameter ($H(t)$) with the {\bf K-}essence scalar field as
\ben
3\frac{\dot a}{a}=3H(t)=-\frac{\ddot\phi}{\dot\phi(1-\dot\phi^{2})},
\label{29}
\een
with the fact that $\dot{a}\neq 0$.

We should pay attention to the value of $\dot{\phi}^2$ in Eq. (\ref{28}). It is obvious that $\dot\phi^{2}< 1$ should
always hold true to get a meaningful signature of the emergent metric Eq. (\ref{28}). Also $\dot\phi^{2}\neq 0$ condition should be hold true to apply the {\bf K-}essence theory. The non-zero value of $\dot{\phi}^2$ is also required to consider $\dot{\phi}^2$ as dark energy density $(\Omega_{DE})$ in units of the critical density, where $\Omega_{Matter} +\Omega_{Radiation} +\Omega_{DE}= 1$ is always true. It is noted that the following works \cite{gm1,gm2,gm3,gm6} support the fact that $\dot{\phi}^2$ is considered as $(\Omega_{DE})$ and therefore should take values between $0<\dot{\phi}^2<1$.

Following the similar mathematical process as \cite{Manna} and using Eqs. (\ref{18}) and (\ref{19}), we can write the components of the Ricci tensors as
\ben
\R_{ii}=\frac{a^{2}}{1-\dot\phi^{2}}\Big[\dot H+H^{2}(3-\dot\phi^{2})\Big]=\frac{a^{2}}{(1-\dot\phi^{2})}\Big[\frac{1}{6}\R(1-\dot\phi^{2})+H^{2}\Big],
\label{30}
\een
and
\ben
\R_{00}=-3\Big[\dot H+H^{2}(1-\dot\phi^{2})\Big]=-\frac{1}{2}(1-\dot\phi^{2})\R+3H^{2},\nonumber\\
\label{31}
\een
where the Ricci scalar is
\ben
\R=-\frac{6}{1-\dot\phi^{2}}\Big[\dot H +H^{2}(2-\dot\phi^{2})\Big].
\label{32}
\een

Now, we assume that the energy-momentum tensor $(\bar{T}_{\mu\nu})$ has the form of an ideal fluid, we can write
\ben
\T_{\mu}^{\nu}&=& diag(\bar{\rho},-\bar{p},-\bar{p},-\bar{p})=(\bar{\rho} +\bar{p})u_{\mu}u^{\nu}-\delta_{\mu}^{\nu} \bar{p}\nonumber\\
\T_{\mu\nu}&=&\G_{\mu\a}\T^{\a}_{\nu},
\label{33}
\een
where $\bar{p}$ is pressure and $\bar{\rho}$ is the matter density of the cosmic fluid in emergent gravity. In the co-moving frame we have $u^{0}=1$ and $u^{\a}=0$ ; $\a= 1, 2, 3$  in the {\bf K}-essence emergent gravity spacetime. The form of Lagrangian Eq. (\ref{8}) gives us the validity of using the perfect fluid model with zero vorticity in {\bf K-}essence theory and also the pressure can be expressed through the energy density only \cite{Vikman,Babichev2}. 

Now we want to evaluate the form of $\bar{\Theta}_{\mu\nu}$. Using Eq. (\ref{16}) and the definition in Eq. (\ref{23}) we have
\ben
\bar{\Theta}_{\mu\nu}&=&\G_{\mu\nu}L(X)-2\T_{\mu\nu}-2\G^{\a\b}\frac{\partial^2 L(X)}{\partial \G^{\mu\nu}\partial \G^{\a\b}}.
\label{34}
\een

The $(00)$ and $(ii)$ component of field equation can be written from (\ref{24}) as:
\ben
F\R_{00}-\frac{1}{2}f(\R,\T)\G_{00}+(\G_{00}\bar{\square}-D_{0}D_{0})F=\frac{1}{2}\T_{00}-f_{\T}\T_{00}-f_{\T}\bar{\Theta}_{00}
\label{35}
\een
and 
\ben
F\R_{ii}-\frac{1}{2}f(\R,\T)\G_{ii}+(\G_{ii}\bar{\square}-D_{i}D_{i})F=\frac{1}{2}\T_{ii}-f_{\T}\T_{ii}-f_{\T}\bar{\Theta}_{ii}
\label{36}
\een

Now using Eqs. (\ref{28}) and (\ref{29}) of emergent geometry, the form of the energy-momentum tensor Eqs. (\ref{33}) and (\ref{34}) in the modified field Eqs. (\ref{35}) and (\ref{36}), we have the following two Friedmann equations in the form of $\tilde{\rho}$ and $\tilde{p}$ as:
\ben
{\tilde{\rho}}&&=\frac{1}{(2 f_{\T}+1) \dot{\phi} \Big(\dot{\phi}^2-1\Big)^3}\Bigg[6 \dot{\phi}^2 \Big(\dot{F} \ddot{\phi}-F \dot{\ddot{\phi}}\Big)-6 \dot{F} \ddot{\phi}+\dot{\phi} \Big(24 F \ddot{\phi}^2+f(\R,\T)+2 f_{T} \Big(\sqrt{1-\dot{\phi}^2}-1\Big)\Big)\nonumber\\
&&+6 F \dot{\ddot{\phi}}+\dot{\phi}^7 \Big(f_{\T} \Big(\sqrt{1-\dot{\phi}^2}+2\Big)-f(\R,\T)\Big)+\dot{\phi}^5 \Big(3 f(\R,\T)+f_{\T} \Big(11 \sqrt{1-\dot{\phi}^2}-6\Big)\Big)\nonumber\\&&
-3 \dot{\phi}^3 \Big(f(\R,\T)+2 f_{\T} \Big(\sqrt{1-\dot{\phi}^2}-1\Big)\Big)+4 f_{\T} \sqrt{1-\dot{\phi}^2} \dot{\phi}^{11}-12 f_{\T} \sqrt{1-\dot{\phi}^2} \dot{\phi}^9\Bigg],
\label{37}
\een
\ben
\tilde{p}=&& \frac{1}{(2 f_{\T}+1) \dot{\phi} \Big(\dot{\phi}^2-1\Big)^3}\Bigg[\dot{\phi} \Big(2 \ddot{F}-8 F \ddot{\phi}^2-f(\R,\T)-2 f_{\T} \sqrt{1-\dot{\phi}^2}+2 f_{\T}\Big)+\dot{\phi}^5 \Big(2 \ddot{F}-3 \Big(f(\R,\T)\nonumber\\&&+2 f_{\T} \Big(\sqrt{1-\dot{\phi}^2}-1\Big)\Big)\Big)+\dot{\phi}^3 \Big(-4 \ddot{F}+3 f(\R,\T)+6 f_{\T} \Big(\sqrt{1-\dot{\phi}^2}-1\Big)\Big)+\dot{\phi}^2 \Big(2 F \dot{\ddot{\phi}}-8 \dot{F} \ddot{\phi}\Big)\nonumber\\&&+ \Big(4 F \ddot{\phi}-2 F \dot{\ddot{\phi}}\Big)+4 \dot{F} \dot{\phi}^4 \ddot{\phi}-4 F \frac{\ddot{\phi}^2}{\dot{\phi}}+\dot{\phi}^7 \Big(f(\R,\T)+2 f_{\T} \Big(\sqrt{1-\dot{\phi}^2}-1\Big)\Big)\Bigg].
\label{38}
\een

It is important to note that the aforementioned $\tilde{\rho}$ and $\tilde{p}$ do not quite align with the values derived from the definition of the emergent energy-momentum tensor (\ref{16}) while considering the perfect fluid energy-momentum tensor (\ref{33}). The analysis of the energy-momentum tensor of a perfect fluid may be conducted within the framework of the {\bf K-}essence geometry. Given that the Lagrangian (\ref{8}) and the emergent energy-momentum tensor (\ref{16}) of the {\bf K-}essence is not directly reliant on the field variable $\phi$, but rather on its time derivative, it is evident that the ideal fluid type energy-momentum tensor may be employed in this particular geometry, as demonstrated by Vikman et al. \cite{Babichev2,Vikman}. In our study, we utilize the ideal fluid emergent energy-momentum tensor (\ref{33}) inside the framework of the $f(\R,\T)$ gravity.

\section{Variations of EoS parameter with time}
In this section, we have discussed the variations of EoS parameter $(\omega)$ with time for different choices of $f(\R,\T)$. The function $f(\R,\T)$ exhibits a range of possible combinations. We adopt the assumption that the dependency in concern may be expressed additively as $f(\R,\T)=f(\R)+f(\T)$. Here, both $\R$ and $\T$ are considered implicit functions of time through the {\bf K}-essence term ($\dot{\phi}^2$), and the function $f(\R,\T)$ is  not explicitly dependent on ($\dot{\phi}^2$) or time. It is important to acknowledge that the energy-momentum tensor, denoted as $\T_{\mu\nu}$, provides information on the energy density and pressure of the cosmos as it evolves across cosmic time. It is essential to remember that the function has the potential to include both cross terms of $\R$ and $\T$, as well as higher order terms of $\T$. These options may have significant significance and may require further investigation in future research. However, now, our intention is to adhere to the most basic options involving curvature ($\R$) and matter ($\T$) such as the additive nature between them.\\

{\bf A. $f(\R,\T)=\R+\l\T$}\\

This is the simplest form of $f(\R,\T)$. This type of function gives us the following results:
$F=\frac{\partial f(\R,\T)}{\partial \R}=1;~
\dot{F}=\frac{\partial F}{\partial t}=0;~
\ddot{F}=\frac{\partial^2 F}{\partial t^2}=0.$

Using Eqs. (\ref{37}) and (\ref{38}) we can evaluate the Friedmann equation for this type of $f(\R,\T)$. Following \cite{gm4}, we choose $\dot{\phi}^2$ as
\ben
\dot{\phi}^2=e^{-\frac{t}{\tau}}\theta(t),
\label{39}
\een
where $\tau$ is a positive constant. 

The above choice of the dark energy density, i.e., kinetic energy of the {\bf K-}essence scalar field appears to be a reasonable one subject to the restriction on $\dot{\phi}^{2}$ ($0<\dot{\phi}^{2}<1$). The modified field Eq.s have been mentioned in {\bf Appendix A} (Eqs. \ref{A1} and \ref{A2}). The suffix $1$ in $\tilde{\r}$ and $\tilde{p}$ has been used to denote the first type of choice of $f(\R)$. 

Solving the field Eqs. (\ref{35}) and (\ref{36}), we can evaluate the EoS parameter from the definition as:
\ben
\omega_{1}=\frac{\tilde{p}_1}{\tilde{\r}_1}=\frac{\o_{N_1}}{\o_{D_1}},
\label{40}
\een
where $\o_{N_1}$ denotes the numerator and $\o_{D_1}$ is the denominator of the EoS parameter respectively. We mentioned the value of $\o_{N_1}$ and $\o_{D_1}$ in {\bf Appendix A} (Eqs. \ref{A3} and  (\ref{A4}).

Now, let us plot this $\omega_{1}$ with time ($t$) for various parametric values of $\tau=1, 2$ (Figs. 1(a) and 1(b)) with different $\lambda (=-0.001,-0.05,-0.10,-0.15)$. 

\begin{figure*}[htp]
  \centering
  \subfigure[$\tau=1$]{\includegraphics[scale=0.6]{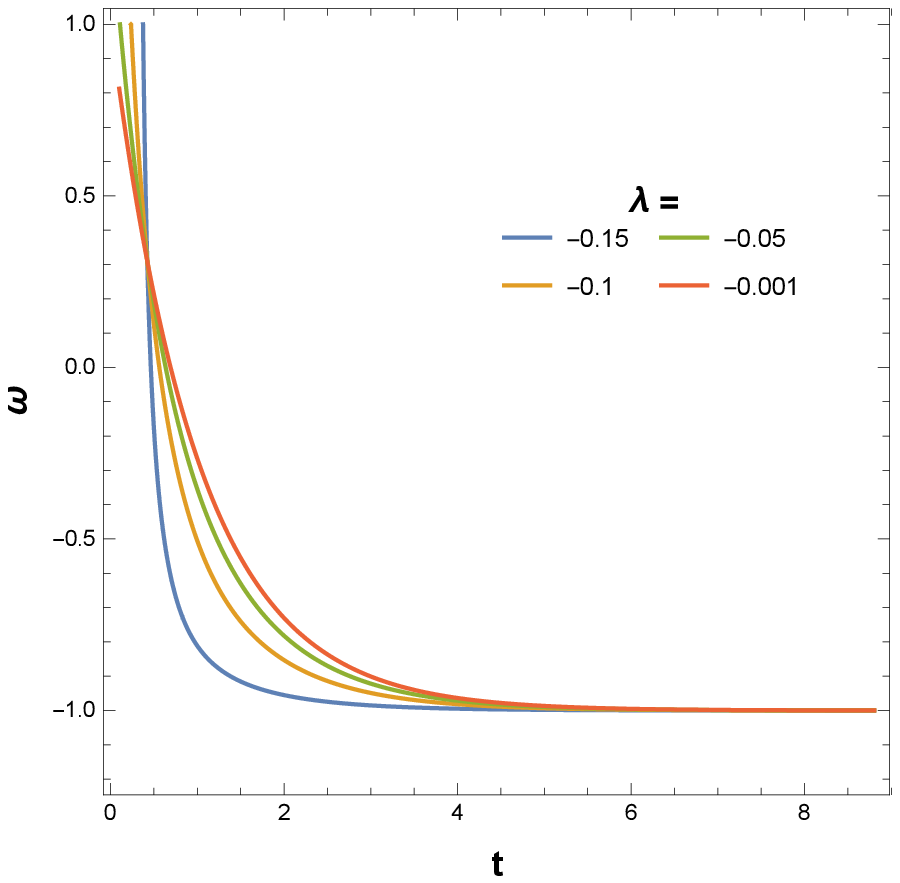}}
  \label{Fig1}
  \quad
  \subfigure[ $\tau=2$]{\includegraphics[scale=0.6]{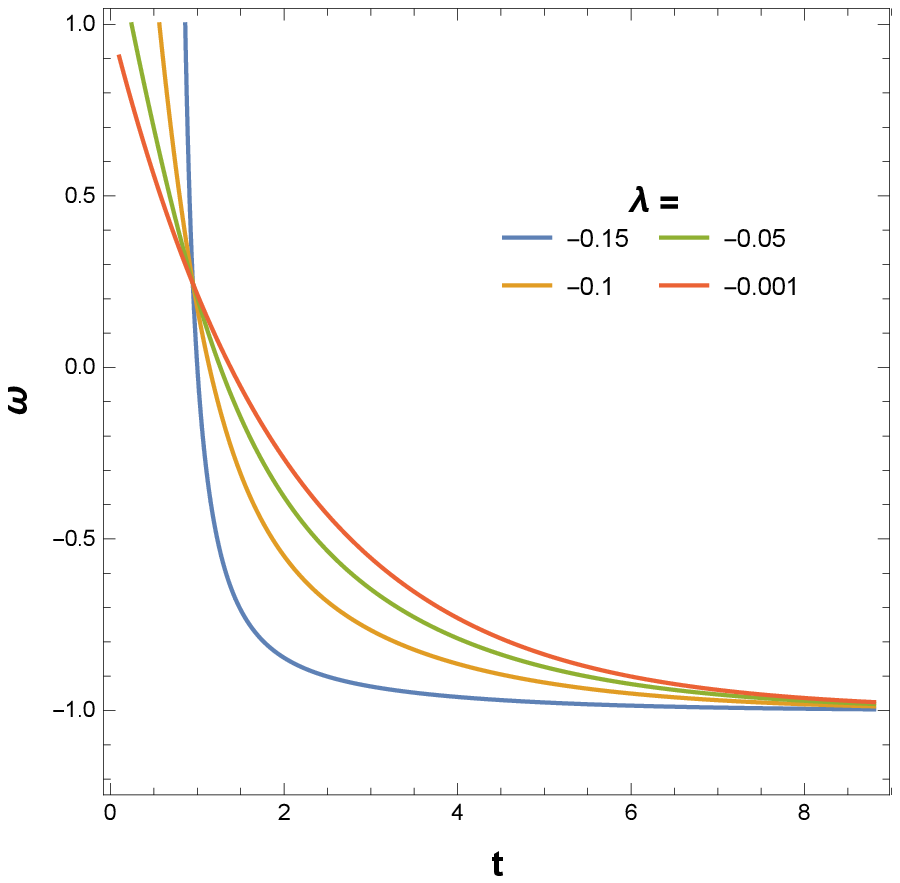}}
  \label{Fig2}
  \caption{Variation of $\omega_1$ with $t$ for $\tau=1,~\&~2$  for different values of $\lambda$ ($=-0.001,-0.05,-0.10,-0.15$)}
\end{figure*}


In Figs. 1(a) and 1(b) we see that the value of $\o$ starts with a positive one and ends up being constant at around $-1$. This curve shows the transition of $\o$ from the radiation-dominated era ($\o=1/3$) to the dark energy-dominated era ($\o=-1$) through the matter-dominated era ($\o=0$). These figures show that the value of $\omega$ decreases with time and takes negative values for each value of $\lambda$, which is satisfying from the observational viewpoint. The observational data \cite{Tripathi} depicts that the value of the EoS parameter should take a value between $-0.95$ to $-1.13$ at the current epoch. Fig. 1(a) represents the variation of the EoS parameter for $\tau=1$ whereas Fig. 1(b) represents the variation of the same with $\tau=2$. In both the figures, we note that $\o$ becomes constant at around $-0.95$ which is the observed value too. It is also observed that for more negative values of $\l$ the steepness of the curve is greater and when the value of $\l$ approaches zero the curve becomes less steep.\\
\vspace{0.3in}

{\bf B. $f(\R,\T)=\R+\alpha \R^2+\l\T$}\\

This special type of $f(\R)$ is known as the Starobinsky type model of modified gravity \cite{moraes2}. Interestingly, Starobinsky modified the Einstein-Hilbert action by considering the second-order term of Ricci scalar with a constant $\alpha$ \cite{Star}, originally formulated in \cite{Star2}. He showed that a cosmological model obtained from the above consideration of $f(R)$ satisfies the cosmological observational tests. This model also forecasts an overproduction of scalars in the very early universe. Considering axially symmetric dissipative dust under geodesic conditions, Sharif et al. \cite{Sharif} investigated the source of the gravitational radiation in the Starobinsky model. Therefore, we take this model into our consideration to check the cosmological scenarios for our case.

For this type of $f(\R)$ we have
\ben
F(\R)=1+2\alpha \R;~~
\dot{F}=2\alpha \dot{\R};~~
\ddot{F}=2\alpha \ddot{\R}.
\label{41}
\een

For this case, the Friedmann equation can be written from the $(00)$ and $(ii)$ components of the field equations (\ref{35}) and (\ref{36}). We can solve those Friedmann equations to get the expression of $\tilde{\r}$ and $\tilde{p}$ in {\bf Appendix A} (Eqs. \ref{A5} and \ref{A6}), which immediately gives us the EoS parameter as 
\ben 
\o_{2}=\frac{\tilde{p}_2}{\tilde{\r_2}}=\frac{\o_{N_2}}{\o_{D_2}},
\label{42}
\een
where $\tilde{\r_2}$ and $\tilde{p}_2$ are energy density and pressure for the second choice of $f(\R,\T)$. The expression of $\o_2$ has been given in {\bf Appendix A} (Eqs. \ref{A7} and \ref{A8}).

The plots of $\o_2$ vs. $t$ has been shown here. We choose two sets of the parameters:\\
I. $\tau=1$, $\l=1$, and $\a$ varies from $1,2,3,4,5$ (Fig. 2(a)).\\
II. $\tau=1$, $\l=5$, $\a=1,2,3,4,5$ (Fig. 2(b)).\\

From these figures, we see that both the curves become constant at $\o_2=-1$ which is consistent with the observational data.


\begin{figure*}[htp]
  \centering
  \subfigure[ $\l=1$ ]{\includegraphics[scale=0.55]{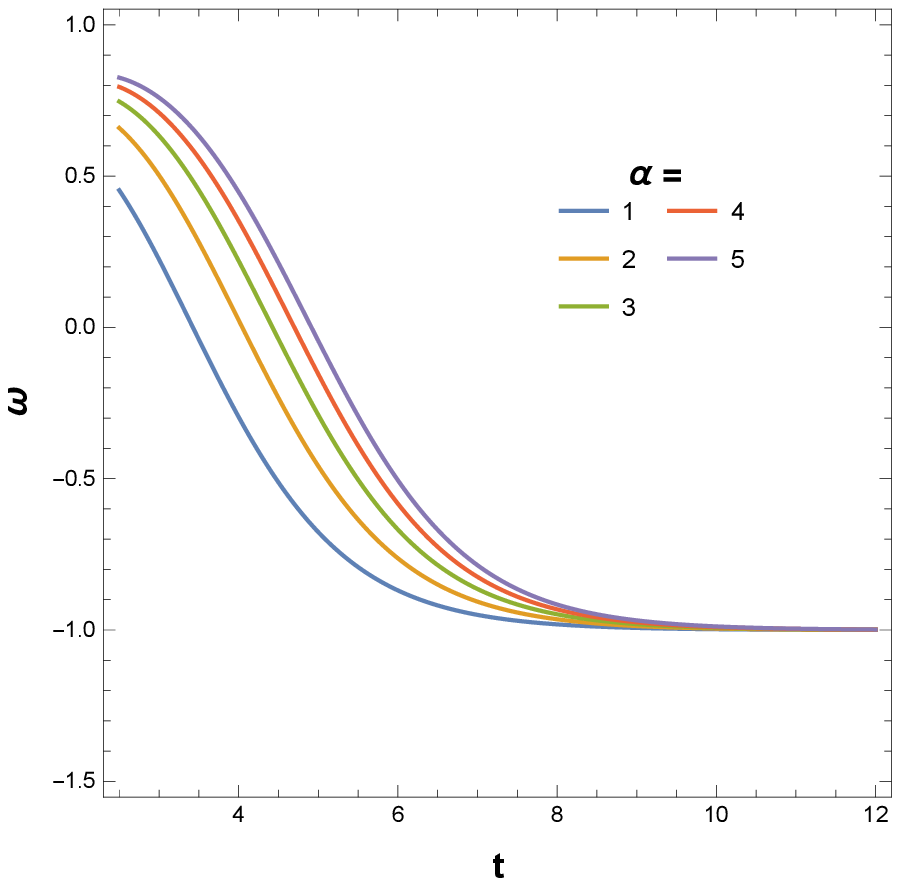}}
  \label{Fig3}
  \quad
  \subfigure[$\l=5$]{\includegraphics[scale=0.55]{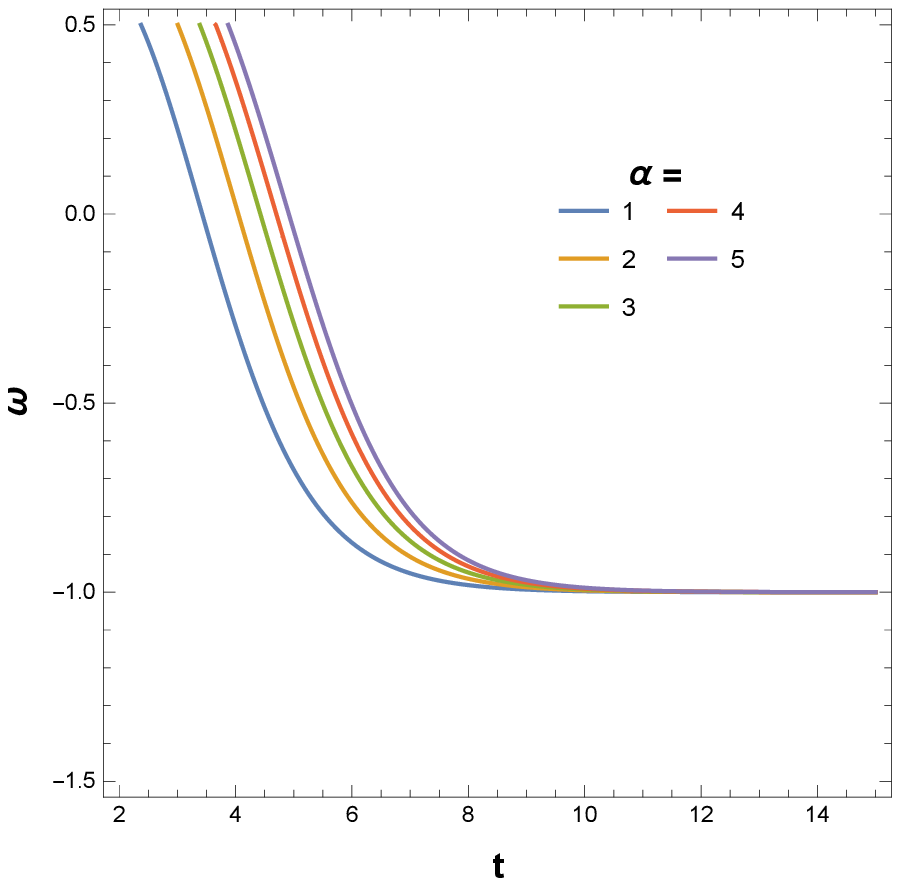}}
  \label{Fig4}
  \caption{Variation of $\omega_2$ with $t$ for $\tau=1$, $\l=1~\&~5$ and different values of $\a$ ($=1,2,3,4,5$)}
\end{figure*}



For this case, we varied the model parameter $\a$ keeping $\l$ fixed. If we look at the time scale we see that for a greater value of $\a$ the transition from radiation ($\o=1/3$) to dark energy-dominated era ($\o=-1$) is delayed. So the greater contribution of the higher order term of Ricci scalar ($\R$) makes the EoS parameter decrease at the late time of evolution.
\vspace{0.3in}

{\bf C. Generalising the choice of \texorpdfstring{$f(\R,\T)$}{TEXT}}\\

Now we want to generalize the form of the function $f(\R,\T)$ and check the validity of each model depending on the integer values of $n$. We take $f(\R,\T)$ as
\ben
f(\R,\T)=\R+\a\R^n+\l \T.
\label{43}
\een
For Eq. (\ref{43}) we get, $F=\frac{\partial f(\R,\T)}{\partial \R}=1+n\a\R^{n-1},~~\dot{F}=\frac{\partial F}{\partial t}=\a n (n-1)\R^{n-2}\dot{\R},\\~\ddot{F}=\frac{\partial^2 F}{\partial t^2}=\a n(n-1)\Big((n-2)\R^{n-3}+\R^{n-2}\ddot{\R}\Big)$ \\

Considering Eqs. (\ref{37}), (\ref{38}) and (\ref{39}), for the above choice of $f(\R,\T)$ (\ref{43}), we have the EoS parameter
\ben 
\o_{3}=\frac{\o_{N_3}}{\o_{D_3}},
\label{44}
\een
where the explicit forms of $\o_{N_3}$ and $\o_{D_3}$ are expressed in (\ref{A9}) and (\ref{A10}) respectively.

We want to check the validity of our model with observational data set \cite{Tripathi} for general choice of $f(\R,\T)$ (\ref{43}) depending on model parameters, which give us some valid plots between $\o$ and time. We have four parameters here which are $\tau,~\a,~\l,~\&~n$. The constraints in choosing these parameters are mentioned below:

\begin{table}
\begin{center}
\resizebox{9cm}{!}{
\begin{tabular}{ |c|c|c| } 
\hline
\text{Parameter} &  \text{Constraint} & \text{Possible Values} \\
\hline
$\tau$ &  \text{can only be positive integer} & $1~\&~10$ \\
$\a$  &\text{can be positive mostly} & $1~\&~10$ \\
  $\l$  & \text{can be any rational number} & $-10,~-0.1,~0.1,~\&~10$ \\
    $n$  & \text{can be positive integers mostly}& $0-10$ \\
  \hline
\end{tabular}}
\end{center}
\caption{Table for possible values of model parameters}
\label{Table1}
\end{table}

These choices of model parameters provide us the opportunity to combine them in various ways to get different graphs of $\o$. With these parameters, we get a $32$ set of parameters which have been mentioned in Table II. From those $32$ sets, we get a physical and viable graph for  $16$ set of parameters only. We have mentioned which set of parameters gives us a valid plot of $\o$ vs. $t$ which actually satisfies the observational data. From Table II, we see that a large value of $\tau$ is not a preferable choice because for large value of $\tau$ graphs do not satisfy the observational data. Now we plot these 16 graphs below:

\begin{table}
\begin{center}
\resizebox{9cm}{!}{
\begin{tabular}{ |c|c|c|c|c|} 
\hline
$\tau$ & $\a$ & $\l$ & $n$ & validity\\
\hline
\multirow{2}{4em}{$1$} &   \multirow{2}{4em}{$1$} &  \multirow{2}{4em}{$-10$} & $0,2,4,6,8,10$ & valid (Plot 9)\\
 & & & $1,3,5,7,9$ & valid (Plot 1)\\
\hline
\multirow{2}{4em}{$10$} &   \multirow{2}{4em}{$1$} &  \multirow{2}{4em}{$-10$} & $0,2,4,6,8,10$ & invalid\\

 & & & $1,3,5,7,9$& invalid\\
 \hline
 \multirow{2}{4em}{$1$} &   \multirow{2}{4em}{$10$} &  \multirow{2}{4em}{$-10$} & $0,2,4,6,8,10$ & valid (Plot 10)\\
 
 & & & $1,3,5,7,9$ & valid (Plot 2)\\
 \hline
 \multirow{2}{4em}{$10$} &   \multirow{2}{4em}{$10$} &  \multirow{2}{4em}{$-10$} & $0,2,4,6,8,10$ & invalid\\

 & & & $1,3,5,7,9$ & invalid\\
\hline
 \multirow{2}{4em}{$1$} &   \multirow{2}{4em}{$1$} &  \multirow{2}{4em}{$-0.1$} & $0,2,4,6,8,10$ & valid (Plot 11)\\

 & & & $1,3,5,7,9$ & valid (Plot.3)\\
\hline
 \multirow{2}{4em}{$10$}  &  \multirow{2}{4em}{$1$} &  \multirow{2}{4em}{$-0.1$} & $0,2,4,6,8,10$ & invalid\\

 & & & $1,3,5,7,9$ & invalid\\
 \hline
  \multirow{2}{4em}{$1$} &   \multirow{2}{4em}{$10$} &  \multirow{2}{4em}{$-0.1$} & $0,2,4,6,8,10$ & valid (Plot 12)\\

 & & & $1,3,5,7,9$ & valid (Plot.4)\\
 \hline
 \multirow{2}{4em}{$10$} &   \multirow{2}{4em}{$10$} &  \multirow{2}{4em}{$-0.1$} & $0,2,4,6,8,10$ & invalid\\

 & & & $1,3,5,7,9$ & invalid\\
\hline
\multirow{2}{4em}{$1$} &   \multirow{2}{4em}{$1$} &  \multirow{2}{4em}{$0.1$} & $0,2,4,6,8,10$ & valid (Plot 13)\\

 & & & $1,3,5,7,9$& valid (Plot 5)\\
\hline
\multirow{2}{4em}{$10$} &   \multirow{2}{4em}{$1$} &  \multirow{2}{4em}{$0.1$} & $0,2,4,6,8,10$ & invalid\\

 & & & $1,3,5,7,9$ & invalid\\
 \hline
  \multirow{2}{4em}{$1$} &   \multirow{2}{4em}{$10$} &  \multirow{2}{4em}{$0.1$} & $0,2,4,6,8,10$ & valid (Plot 14)\\

 & & & $1,3,5,7,9$ & valid (Plot 6)\\
 \hline
\multirow{2}{4em}{$10$} &   \multirow{2}{4em}{$10$} &  \multirow{2}{4em}{$0.1$} & $0,2,4,6,8,10$ & invalid\\

 & & & $1,3,5,7,9$ & invalid\\
 \hline
 \multirow{2}{4em}{$1$} &   \multirow{2}{4em}{$1$} &  \multirow{2}{4em}{$10$} & $0,2,4,6,8,10$ & valid (Plot 15)\\

 & & & $1,3,5,7,9$ & valid (Plot.7)\\
 \hline
 \multirow{2}{4em}{$10$} &   \multirow{2}{4em}{$1$} &  \multirow{2}{4em}{$10$} & $0,2,4,6,8,10$  & invalid\\

 & & & $1,3,5,7,9$ & invalid\\
 \hline
 \multirow{2}{4em}{$1$}  &  \multirow{2}{4em}{$10$} &  \multirow{2}{4em}{$10$} & $0,2,4,6,8,10$ & valid (Plot 16)\\

 & & & $1,3,5,7,9$ & valid (Plot 8)\\
 \hline
 \multirow{2}{4em}{$10$} &   \multirow{2}{4em}{$10$} &  \multirow{2}{4em}{$10$} & $0,2,4,6,8,10$ & invalid\\

 & & & $1,3,5,7,9$& invalid\\
 \hline
\end{tabular}}
\end{center}
\caption{Table for a possible set of model parameters which are valid or invalid based on plots.}
\label{Table2}
 \end{table}

\begin{figure}
 \begin{center}
    \subfigure{
            \label{fig:1.1}
            \includegraphics[width=0.4\textwidth]{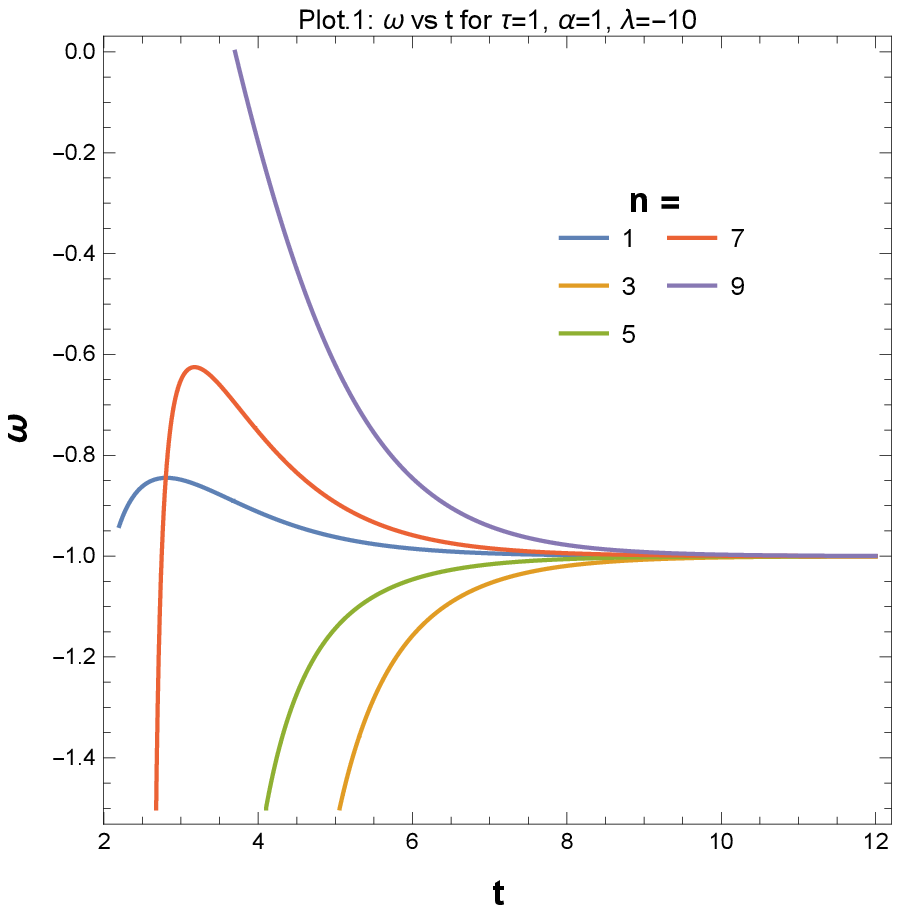}
        }
        \subfigure{
           \label{fig:1.2}
           \includegraphics[width=0.4\textwidth]{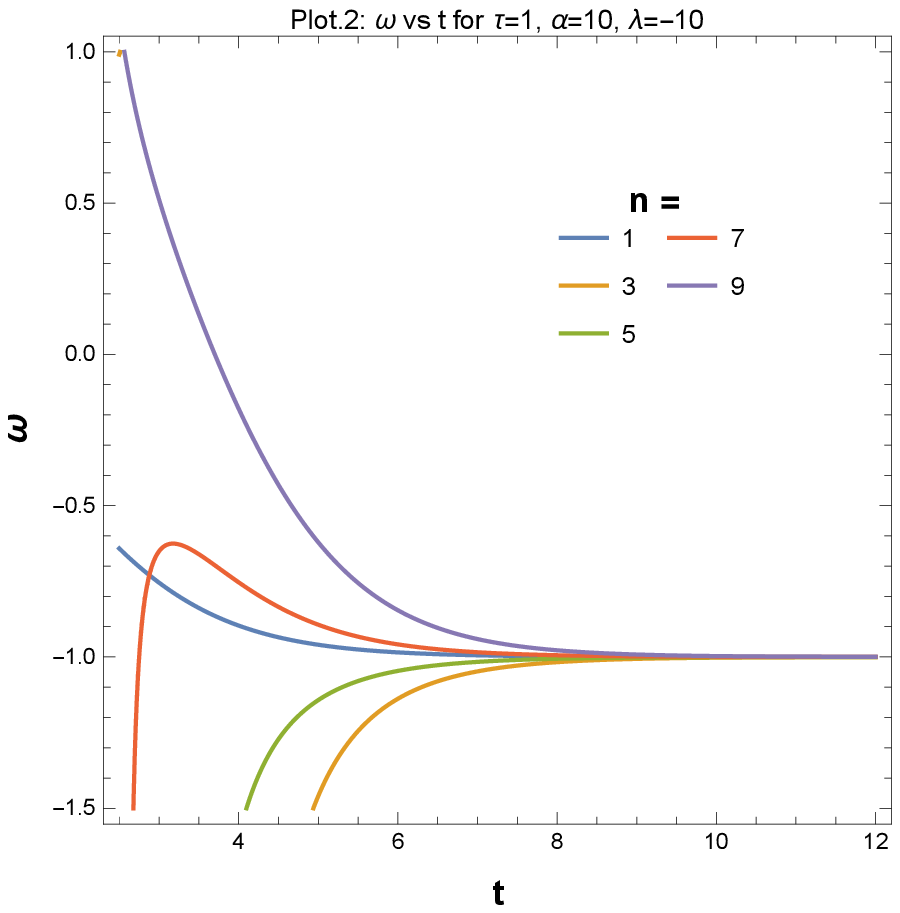}
        }
        \subfigure{
            \label{fig:1.3}
            \includegraphics[width=0.4\textwidth]{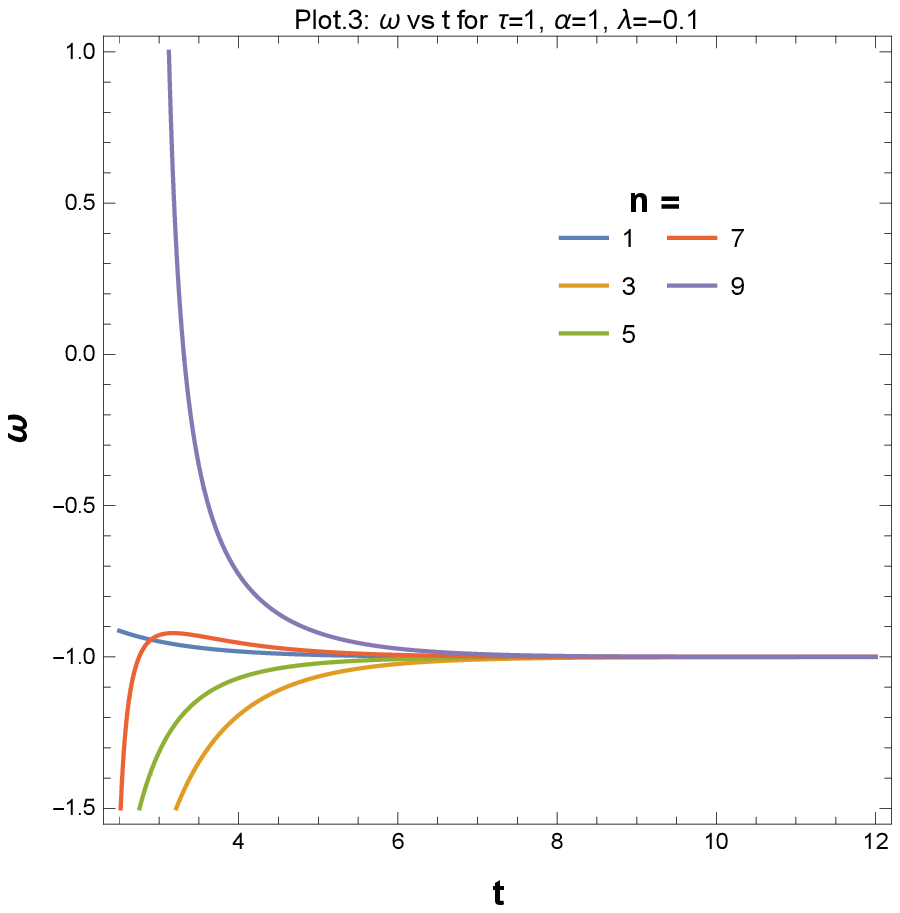}
        }
        \subfigure{
            \label{fig:1.4}
            \includegraphics[width=0.4\textwidth]{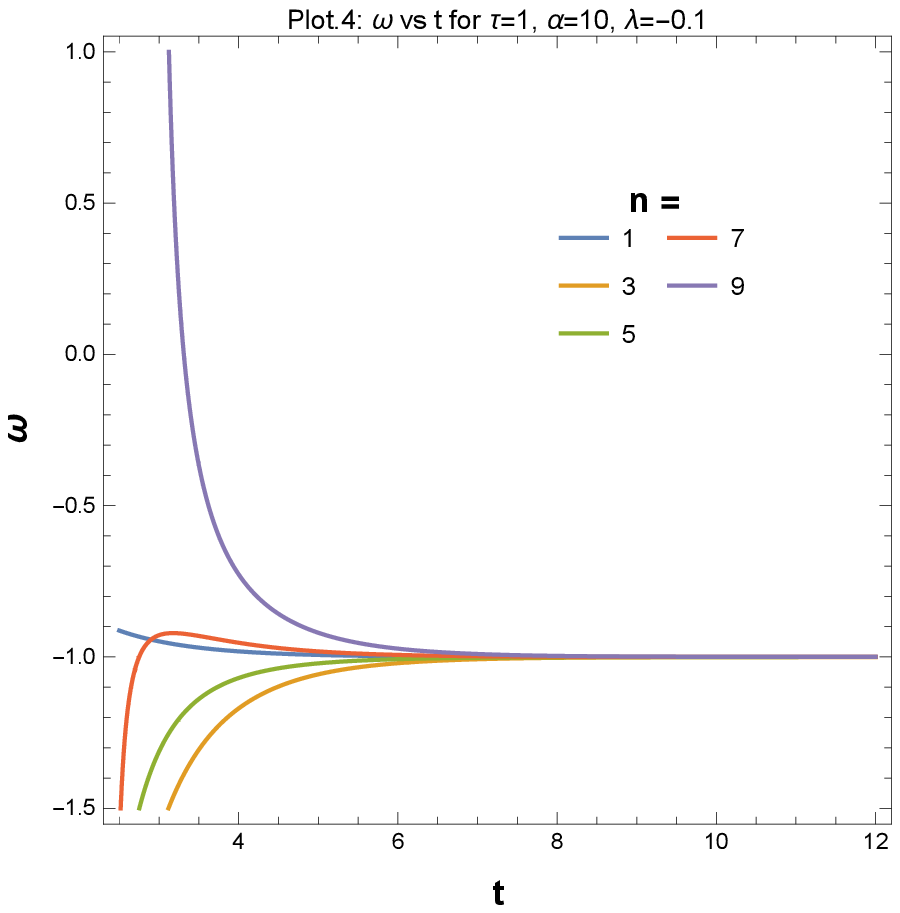}
        }
     \end{center}
     \caption{ 
       Variation of $\o$ with $t$ for different model parameters for odd $n$   
     }
   \label{fig:subfigures1}
\end{figure}

\begin{figure}
     \begin{center}

        \subfigure{
            \label{fig:1.5}
            \includegraphics[width=0.4\textwidth]{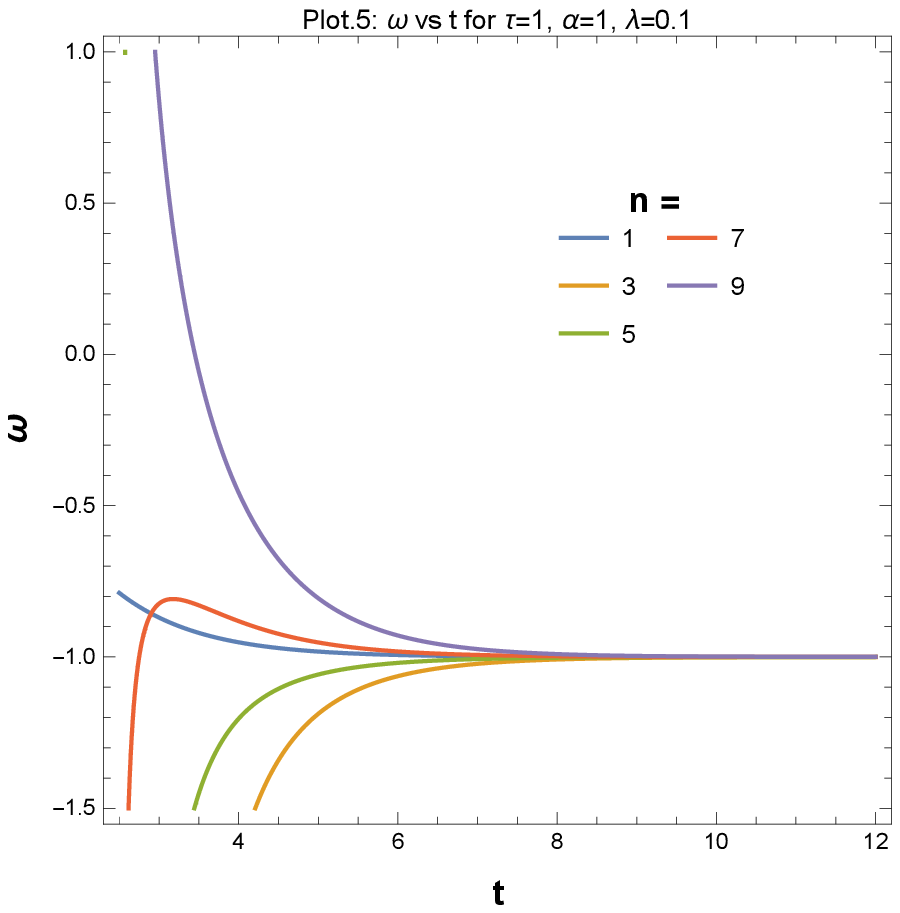}
        }
        \subfigure{
           \label{fig:1.6}
           \includegraphics[width=0.4\textwidth]{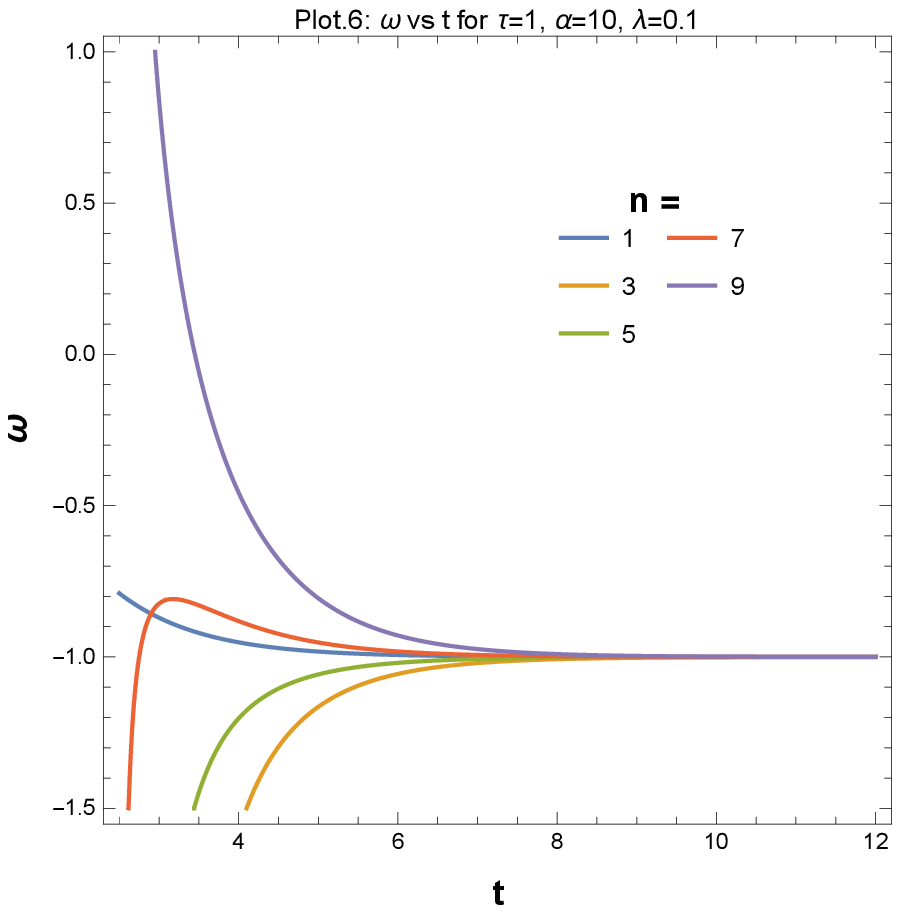}
        }
        \subfigure{
            \label{fig:1.7}
            \includegraphics[width=0.4\textwidth]{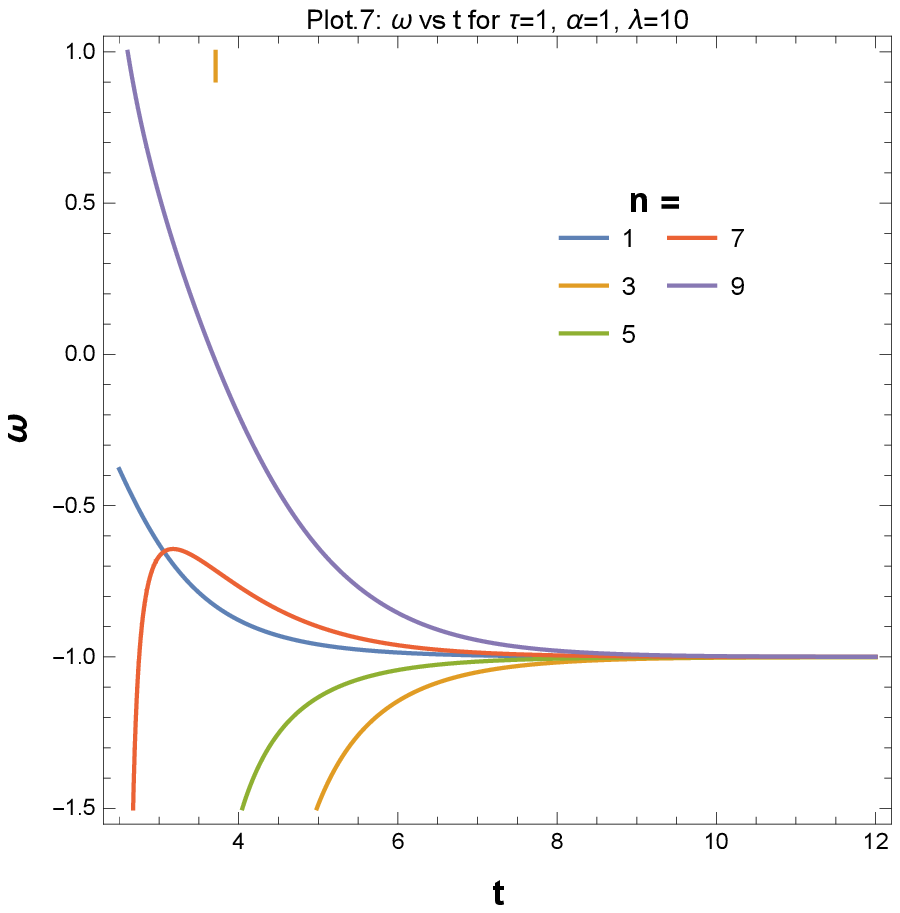}
        }
        \subfigure{
            \label{fig:1.8}
            \includegraphics[width=0.4\textwidth]{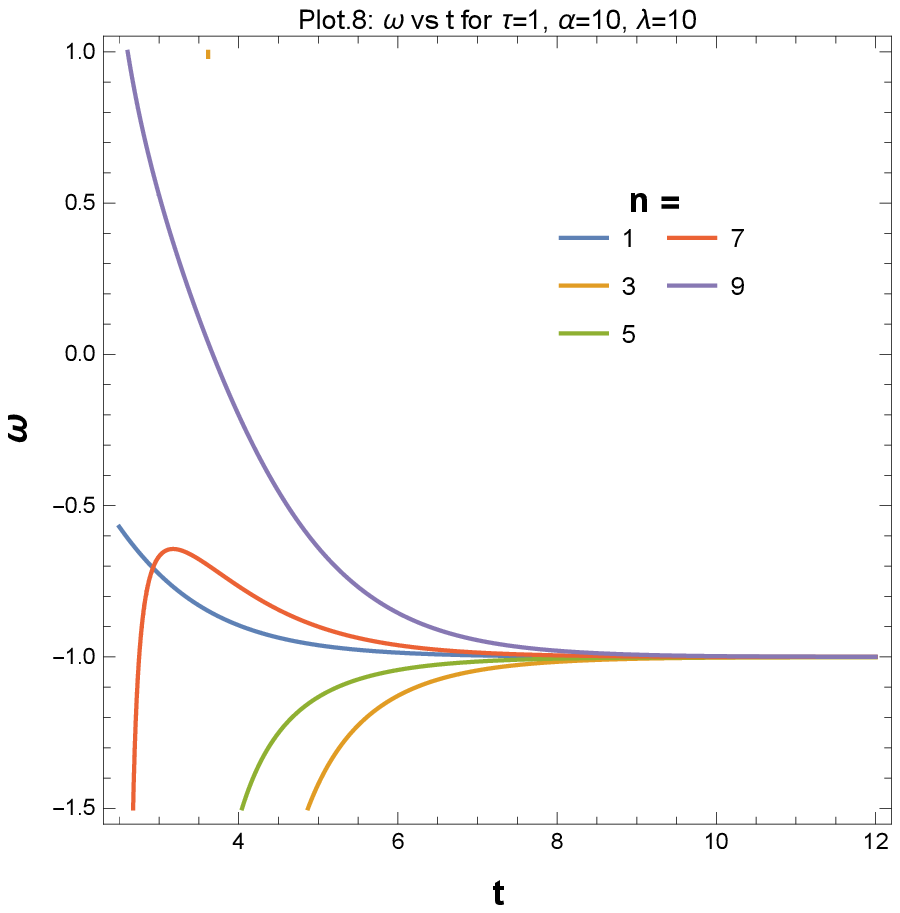}
        }
    \end{center}
    \caption{ 
         Variation of $\o$ with $t$ for different model parameters for odd $n$
     }
   \label{fig:subfigures2}
\end{figure}
 
\begin{figure}
     \begin{center}

        \subfigure{
            \label{fig:1.9}
            \includegraphics[width=0.4\textwidth]{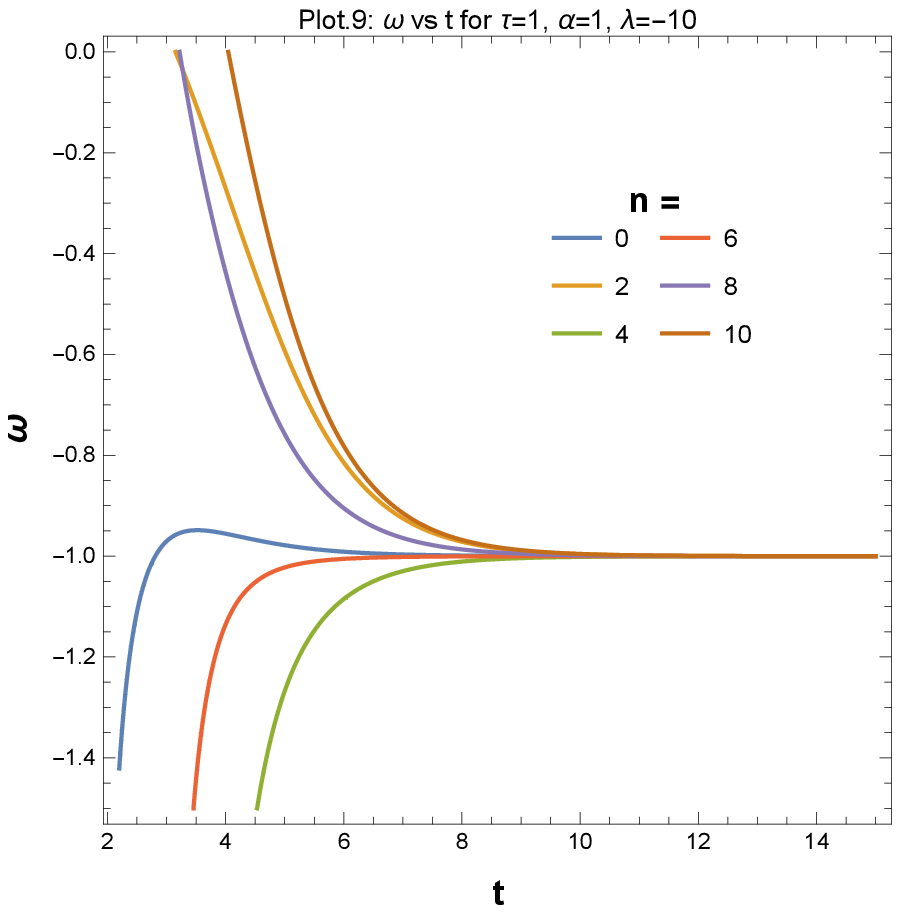}
        }
        \subfigure{
           \label{fig:1.10}
           \includegraphics[width=0.4\textwidth]{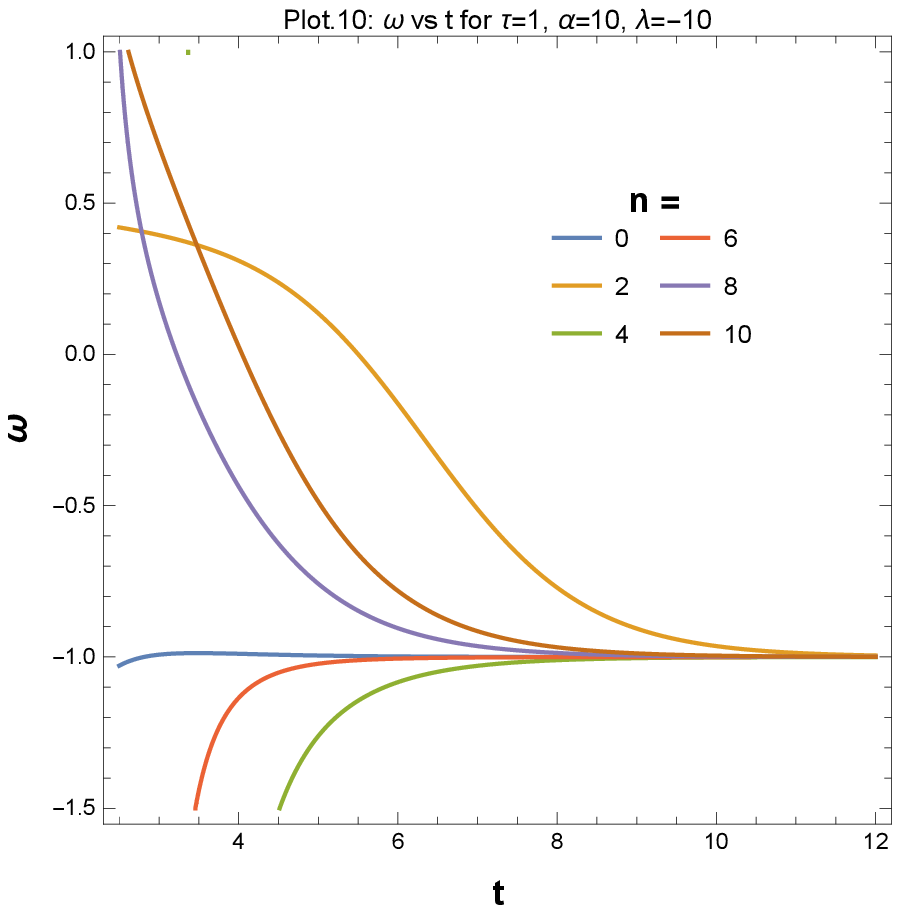}
        }
        \subfigure{
            \label{fig:1.11}
            \includegraphics[width=0.4\textwidth]{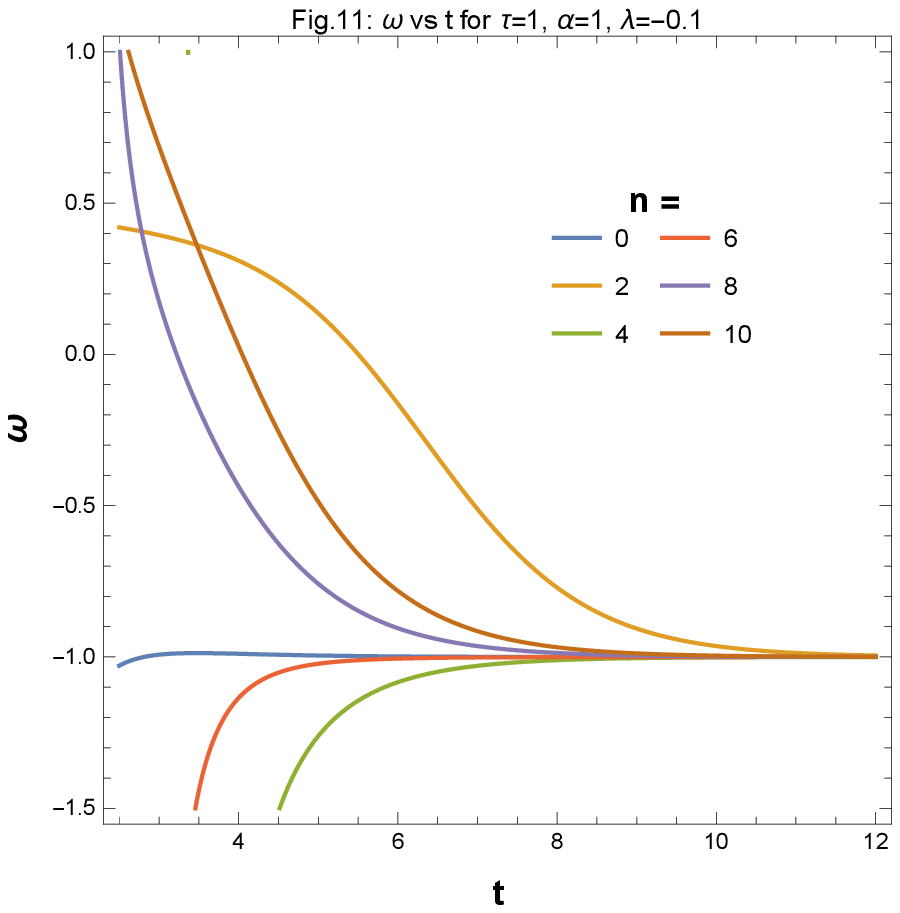}
        }
        \subfigure{
            \label{fig:1.12}
            \includegraphics[width=0.4\textwidth]{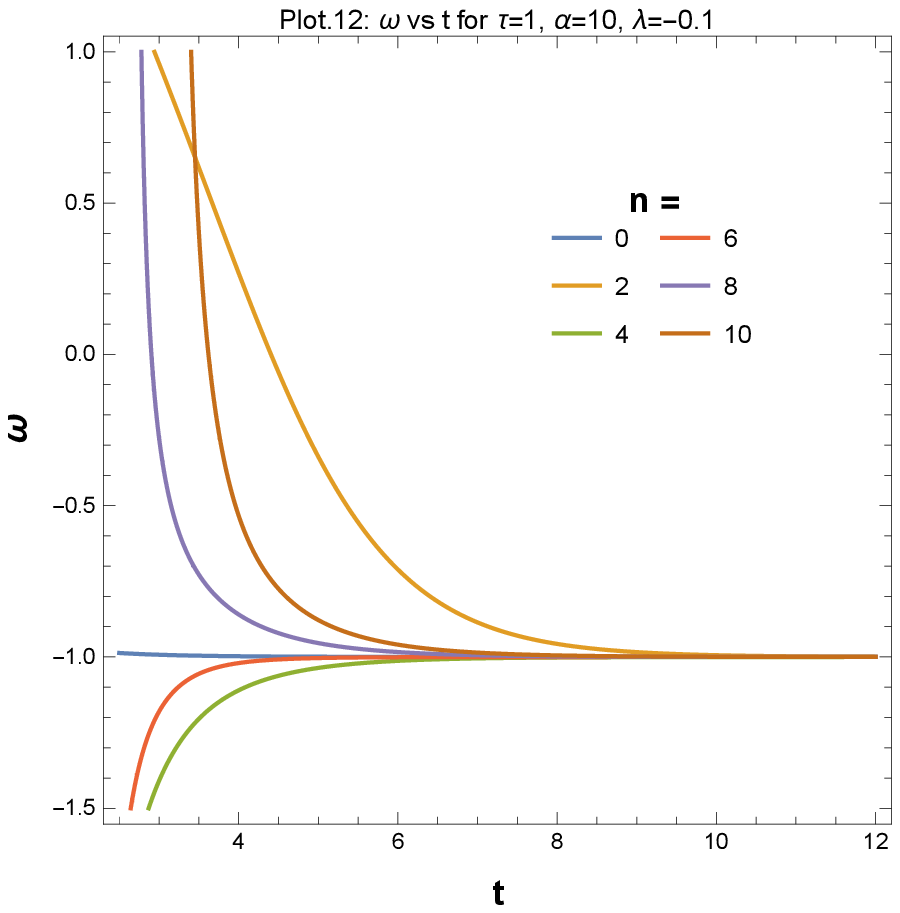}
        }
    \end{center}
    \caption{  
       Variation of $\o$ with $t$ for different model parameters for even $n$   
     }
   \label{fig:subfigures3}
\end{figure}

\begin{figure}
     \begin{center}

        \subfigure{
            \label{fig:1.13}
            \includegraphics[width=0.4\textwidth]{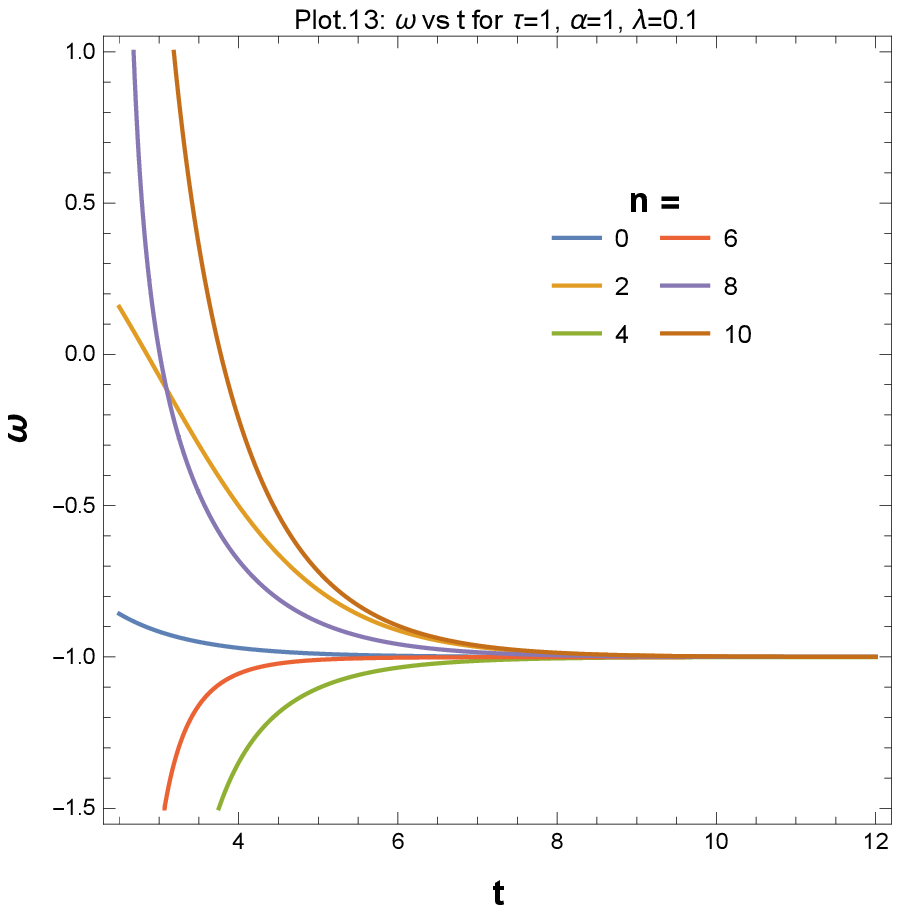}
        }
        \subfigure{
           \label{fig:1.14}
           \includegraphics[width=0.4\textwidth]{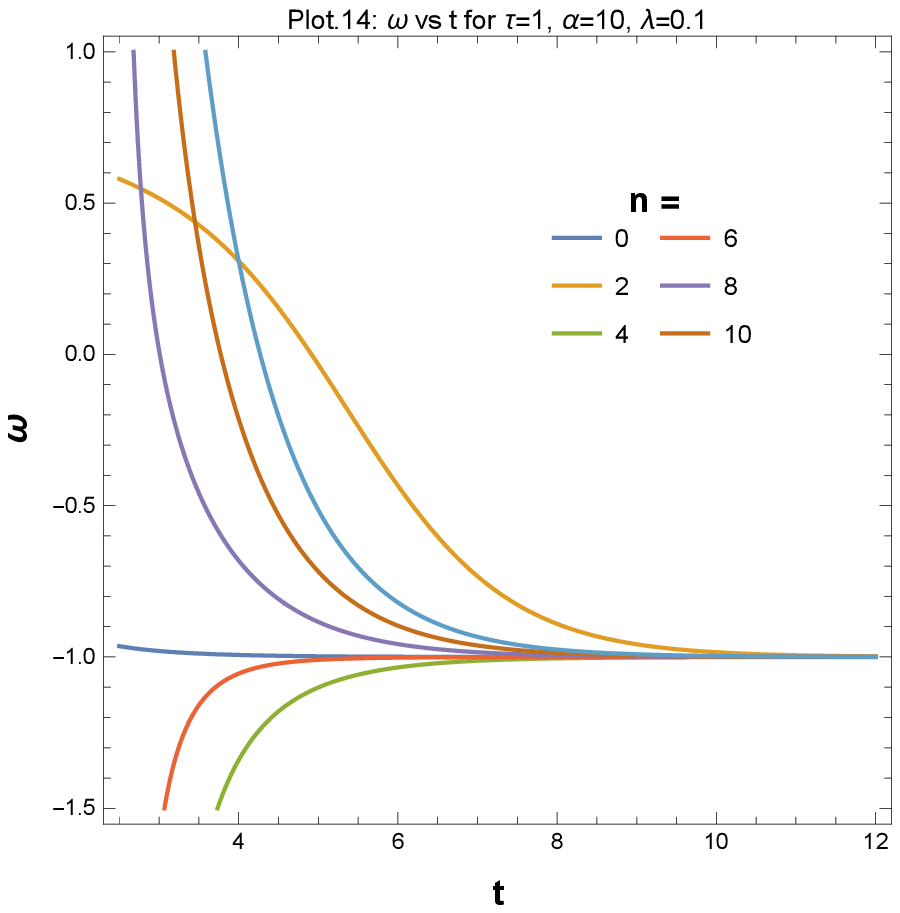}
        }
        \subfigure{
            \label{fig:1.15}
            \includegraphics[width=0.4\textwidth]{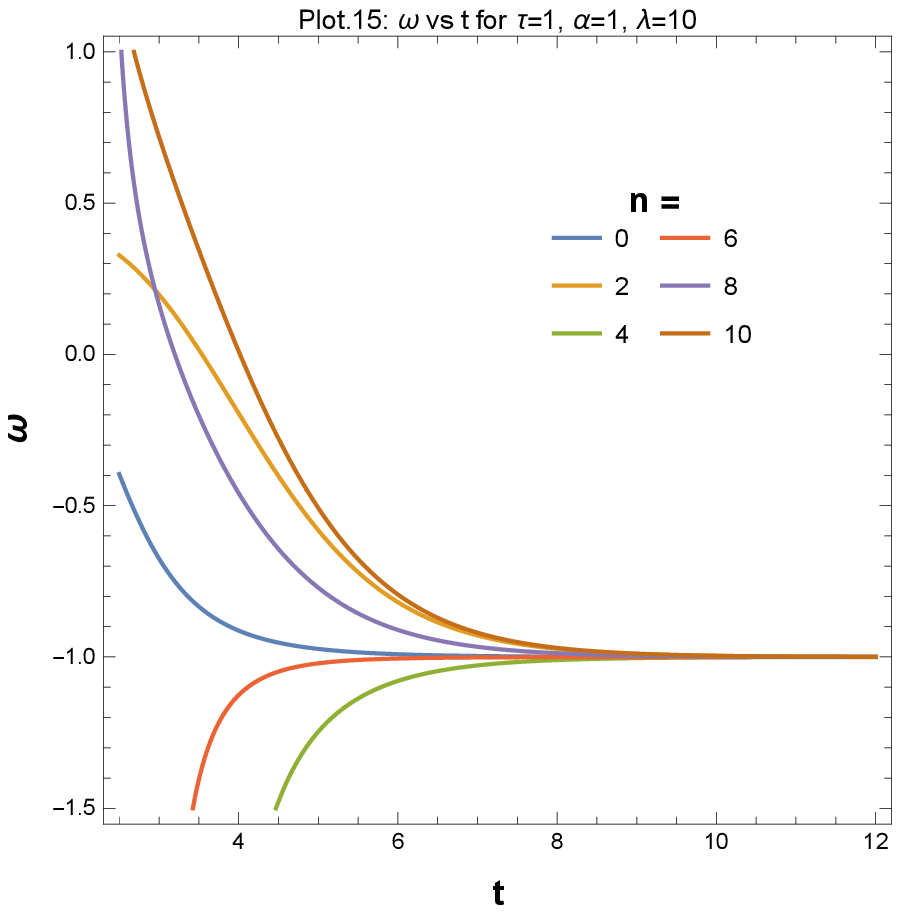}
        }
        \subfigure{
            \label{fig:1.16}
            \includegraphics[width=0.4\textwidth]{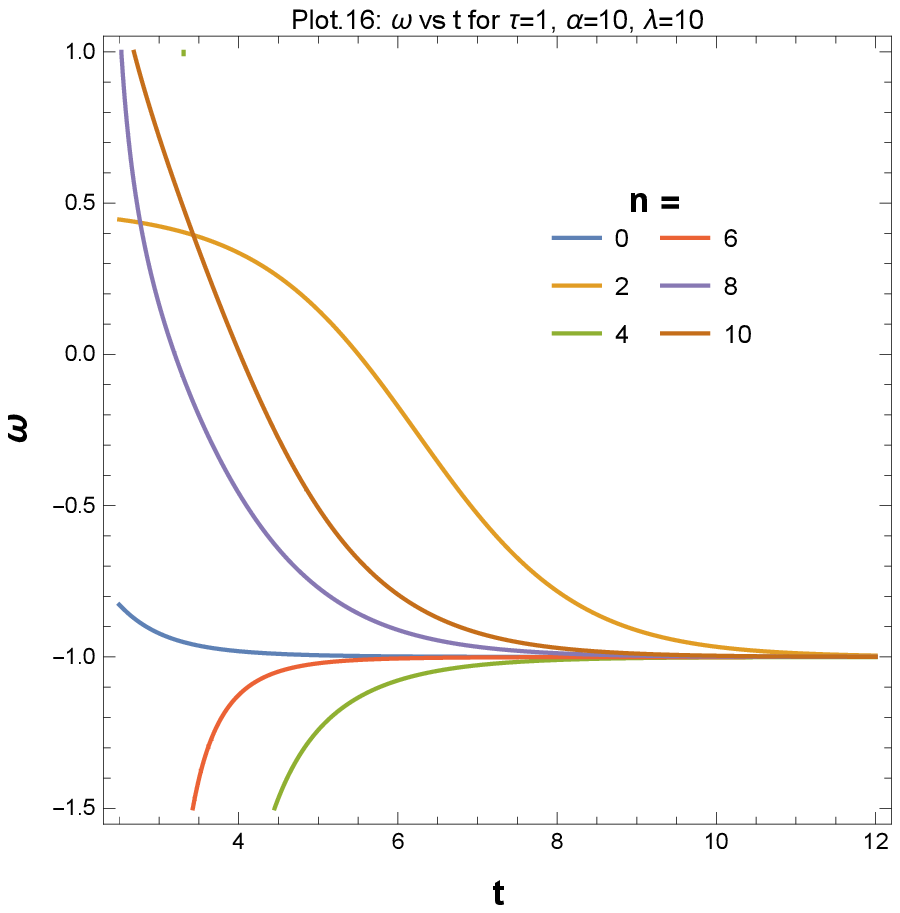}
        }
    \end{center}
    \caption{ 
         Variation of $\o$ with $t$ for different model parameters for even $n$
     }
   \label{fig:subfigures4}
\end{figure}

 Analyzing every graph (Figs. 3,4,5 and 6) shown above we note that in each case the observational data for late time acceleration are satisfied by the lower ($n$=0,~1,~2) and higher ($n$=8,~9,~10) value of $n$. The mid-value of $n$(=3,~4,~5,~6 and 7) produces curves that start from the negative value of $\o$. When the power of $\R$ varies in lower order the curvature is small and when the power of $\R$ is high the curvature varies extensively where $\R$ denotes the curvature in the action. Now for both the lower ($n$=0,~1,~2) and higher ($n$=8,~9,~10) curvatures, we get the observation-satisfied result at the present epoch. Generally, high curvature indicates that the effect of gravity is strong which may be a region around dense objects. For $n=7$ shows a peculiar variation of EoS parameter for some plots, i.e., it starts from a large negative value, then crosses the $-1$ value, and goes upward, then again it becomes constant at $-1$. The mid-value of $n$, i.e., $n$=3,~4,~5,~6, and 7 shows the EoS parameter starts from high negative to $-1$, which is the present value of the EoS parameter of the universe. This indicates that the early universe may be dark energy dominated. This may justify the kinetically driven inflation, i.e. {\it K-inflation} scenario \cite{Picon3}, which means the kinetically driven inflation rolls gracefully from a high-curvature starting phase to a low-curvature phase and may eventually leave inflation to become radiation-dominated. So we may conclude that our model is valid for the entire range of the universe which means starting from the early (ultra-relativistic era) to the present epoch (dark energy-dominated era).  It should be noted that in the above pictures, the time denotes cosmic time and has not been scaled, and $t \to 0$ on this axis indicates the early universe.\\
 \vspace{0.3in}

{\bf D. Observational Verification of the models at the present epoch.}\\

In this sub-section, we compare our results for different choices of $f(\R,\T)$ with the observational data sets \cite{Tripathi}.

\begin{table*}
\begin{center}
\resizebox{11cm}{!}{
\begin{tabular}{ |c|c|c|c|c| } 
\hline
$\tau$ &  $\l$ & $t$ & $\omega_1$ ($3\sigma$ confidence)\cite{Tripathi}  & Observation \cite{Tripathi}\\
\hline
 \multirow{4}{4em}{$1$} & $-0.15$ & 1.91077 & \multirow{4}{4em}{$-0.95\geq\omega\geq-1.13$} & \multirow{4}{4em}{SNIa+ BAO+ H(z)} \\ 
 
  & $-0.1$ & 3.02237  & & \\
  & $-0.01$ & 3.64823 & & \\
  & $-0.001$ & 3.68496 & & \\ 
   
  \hline
 \multirow{4}{4em}{$2$} & $-0.15$ & 3.57211 & \multirow{4}{4em}{$-0.95\geq\omega\geq-1.13$} & \multirow{4}{4em}{SNIa+ BAO+ H(z)} \\ 
 
  & $-0.10$ & 5.98057  & & \\
  & $-0.01$ & 7.29301  & & \\
  & $-0.001$ & 7.36959  & & \\ 
  \hline
 \end{tabular}}
\end{center}
\caption{Table for observational verification of the model for $f(\R,\T)=\R+\lambda \T$ for $\tau=1,2$ and $\l=-0.10,-0.11,-0.12,-0.13, -0.14,-0.15$.}
\label{Table3}
 \end{table*}
 
\begin{table*}
\begin{center}
\resizebox{11cm}{!}{
\begin{tabular}{ |c|c|c|c|c|c| } 
\hline
$\tau$ & $\l$ & $\a$ & $t$ & $\omega_1$ ($3\sigma$ confidence)\cite{Tripathi}  & Observation \cite{Tripathi}\\
\hline
 \multirow{5}{4em}{$1$} & $\multirow{5}{4em}{$1$}$ &$1$ & 6.9989215500337565 & \multirow{5}{4em}{$-0.95\geq\omega\geq-1.13$} & \multirow{5}{4em}{SNIa+ BAO+ H(z)} \\ 
 
  & &$2$ & 7.648323690317918 & & \\
  & &$3$ & 8.038769314737097 & & \\
  & &$4$ & 8.318855798895852  & &\\ 
  & &$5$ & 8.53741401857794 & &\\ 
  \hline
 \multirow{5}{4em}{$1$} & \multirow{5}{4em}{$5$}& $1$ & $6.9989215500337565$ & \multirow{5}{4em}{$-0.95\geq\omega\geq-1.13$} & \multirow{5}{4em}{SNIa+ BAO+ H(z)} \\ 
 
  & &$2$ &  7.648323690317918 & & \\
  & &$3$ & 8.038769314737097  & & \\
  & &$4$ & 8.318855798895852  & &\\ 
  & &$5$ &  8.53741401857794 & &\\ 
  \hline
 \end{tabular}}
\end{center}
\caption{Table for observational verification of the model for $f(\R,\T)=\R+\a \R^2 +\lambda\T$ for $\tau=1$, $\l=1,5$ and $\a=1,2,3,4,5$.}
\label{Table4}
 \end{table*}

Tables III and IV indicate how well our model matches to the observational data as available in Ref. \cite{Tripathi}. We used the EoS parameter data from the combined observations of SNIa, BAO, and H(z), which indicates that for the current accelerating universe, the EoS parameter should have a value in the range $-0.95\geq\omega\geq-1.13$. We found that for certain parameter selections, we may attain a time range in which the above-mentioned value of $\omega$ is satisfied. So, for the above two situations, we can state unequivocally that our model is very much compatible with the observable evidence at the current epoch. In Planck 2015-XIII results \cite{Planck2015XIII}(p.p-39) the authors mentioned that, due to a significant geometrical degeneracy, even for spatially flat models, the CMB temperature measurements alone cannot accurately constrain $\o$. The researchers employed the camb implementation of the ``parameterized post-Friedmann" (PPF) framework to examine the presence of temporal variation in the parameter $\omega$. The objective of this framework is to restore the characteristics shown by canonical scalar field cosmologies that are minimally coupled to gravity in cases where $\omega\geq -1$, while also providing a reliable approximation for scenarios where $\omega\simeq-1$. In the aforementioned scenarios, the velocity of sound is equivalent to the velocity of light, resulting in a significant reduction in the concentration of dark energy inside the observable universe. One of the benefits associated with using the PPF formalism is the ability to investigate the phantom domain, denoted as $\omega< -1$, which encompasses transitions over the ``phantom barrier" at $\omega= -1$.
If only the Plank result is considered the value of EoS parameter comes out to be $-1.54^{+0.62}_{-0.50}$. which has a $2\sigma$ shift into the phantom region. This value has been observationally modified in their paper by combining the Planck temperature+polarization likelihood together with the BAO, JLA,
and $H_0$ data as$-1.019^{+0.075}_{-0.080}$. The authors studied the case of varying EoS parameters by considering the first-order Taylor series expansion in the scale factor (a) as $\o=\o_0+(1-a)\o_a$ and matched their ansatz with the observation. Through this research, it is clearly shown that our theoretical model also meets the values of EoS parameters under CMB data. It should be noted that our approach is entirely based on a theoretical model that prescribes the numerical value of the EoS parameter of the universe's current epoch (rather than an ansatz) that meets the empirical data for a certain time period.  Also, we have not characterized other cosmological parameters like the Hubble parameter ($H_0$), luminosity distance, redshift, and so on in this study. As a result, in our scenario, the CMB analysis or the $\chi^2$ analysis is not required for the theoretical examination of the current framework.

\section{Conclusion and Discussion}
In a nutshell, in this paper, we have tried to formulate the $f(\R,\T)$ gravity theory in a non-standard theory (i.e. {\bf K-}essence theory). Starting with a brief review of the {\bf K-}essence theory for a special type of non-canonical Lagrangian (i.e. DBI type), we entered into our work, which is to establish a different form of the well-known $f(\R,\T)$ gravity theory by the metric formalism. Taking the action of the modified gravity in the context of {\bf K-}essence geometry and following \cite{Harko}, we have presented the modified field equations of our case (\ref{24}). One can easily find that the field equations of our case is different from the standard field equations of $f(R,T)$ gravity \cite{Sahoo3,Harko} due to the presence of the {\bf K-}essence scalar field. 

Not only that, it is also crystal clear that the field equations of usual $f(R,T)$ gravity can be recovered from (\ref{24}), if we consider the {\bf K-}essence scalar field ($\phi$) terms to be zero and the Lagrangian as $\La_{m}=p$. Here we would like to mention that, according to Refs.\cite{gm1,gm2,gm3,Manna,gm6}, the kinetic energy of the {\bf K-}essence scalar field ($\dot{\phi}^2$), considered as the dark energy density (in the unit of critical density) of the present universe. Similar to the works~\cite{Sahoo3} and \cite{Harko}, the covariant derivative of the energy-momentum tensor has also been calculated for our case which, for obvious reasons, turns out to be of different from \cite{Sahoo3} and \cite{Harko}. We get some extra terms related to the scalar field ($\phi$), which may reveal some more astonishing features of this theory. These extra terms appeared due to interactions of the usual gravity with the {\bf K-}essence scalar field. 

We have solved the field equations by considering the background metric as the flat FLRW type and computed the Friedmann equations of our theory for the general form of $f(\R,\T)$. We also have established a new definition of $\bar{\Theta}_{\mu\nu}$ in (\ref{34}), which is not so straightforward as obtained in \cite{Sahoo3} and \cite{Harko}. As the Lagrangian for our case has an implicit dependence on the scalar field $\phi$ of the {\bf K-}essence geometry, specifically it has explicit dependence on the derivative of $\phi$, some extra features are included in it automatically. In this process, we have used some important relations, which connect the usual geometry with the {\bf K-}essence geometry already showed in \cite{gm1,gm2}.

The next part of our work includes the evaluation of the EoS parameter by solving the Friedmann equations obtained in (\ref{37}) and (\ref{38}), where we have considered the general form of $f(\R,\T)$. We get far different results from what has been achieved by Sahoo et al. \cite{Sahoo3} in their Eqs. (13) and (14).  Following \cite{gm4}, we have expressed $\dot{\phi}^2$ as an exponential function of time (\ref{39}) which satisfy the conditions for $\dot{\phi}^{2}~(0<\dot{\phi}^{2}<1)$.  Next, we have found the variation of the EoS parameter by choosing different forms of $f(\R,\T)$. Particularly, we have taken three choices of $f(\R,\T)$, of which the first one is the simple case, i.e., $f(\R,\T)=\R+\lambda\T$ (Section: IV, Case A), the second one is of Starobinsky type $f(\R)$, i.e., $f(\R,\T)=\R+\alpha\R^2+\lambda\T$ (Section: IV, Case B) and third one is more general form, i.e., $f(\R,\T)=\R+\a\R^n+\l \T$ (Section: IV, Case C).

It is a well-known fact, that for our present universe, which has a positive acceleration, the value of the EoS parameter is $\omega\leq-1$. So to check the efficacy of our results we have plotted the variation of this EoS parameter obtained in this paper in Section IV. For each consideration of $f(\R,\T)$, we first set the required parameter values and then we draw the graph of $\omega$ vs. $t$. In Case C, we analyze a generic form of $f(\R,\T)$ and demonstrate through the plots that our model is consistent with the empirical data for particular model parameter choices.
Then in Section IVD, we matched the obtained graphs for first two choices of $f(\R,\T)$ in Table III and Table IV with the observational data taken from \cite{Tripathi}. Note that in these graphs we have taken $t\rightarrow 0$ as the early time of the universe. It is also mentioned that up to Eq. (\ref{25}), we have constructed a generalized non-canonical $f(\R,\T)$ theory on the basis of specific DBI type non-standard action in an emergent {\bf K-}essence geometry. The next portion is a special case study of this theory under the consideration of the relation between the Hubble parameter and the {\bf K}-essence scalar field where we have assumed the background gravitational metric to be flat FLRW type.

From a totally different background, we establish the $f(\R,\T)$ theory of modified gravity with dark energy consideration in a new packet. The achievement of this work lies in the fact that without considering the cosmological constant model we are able to produce a theory that supports the results of present observational data. It is also noted that the {\bf K-}essence theory is observationally tested through the Planck collaboration results: 2015-XIV \cite{planck0}, 2018-VI \cite{planck2}.
From the results and graphs obtained here, we see that there are some choices of model parameters ($\lambda, \tau, \alpha$) for which we get $\omega\leq -1$, which satisfy the observational data to a large extent.

The mid-value of $n$, or $n$=3,~4,~5,~6,~7, indicates that the EoS parameter begins from high negative to $-1$, which is the current value of the EoS parameter of the universe, for the general case of $f(\R,\T)$ (Case C). This suggests that dark energy could have been dominant in the early cosmos. This may support the kinetically driven inflation, often known as the {\it K-inflation} scenario \cite{Picon3}, which smoothly transitions from a high-curvature initial phase to a low-curvature intermediate phase before leaving inflation to become radiation-dominated. We may thus draw the conclusion that our model holds true for the evolution of the entire phases of the universe, beginning from the early era and ending in the present. It has also been noted that this theory may be used not just in the context of dark energy, but also from a purely gravitational standpoint \cite{gm4,gm7,Ray}, since the existence of dark energy in the context of cosmology is still a matter of debate \cite{Nielsen}.  Numerous studies have been conducted in the area of cosmology, focusing on modified gravity theories such as $f(R,T)$ gravity, {\bf K-}essence, and DBI-action. These theories have been investigated individually or in combination with each other, primarily aiming to explain the cosmic phenomena of late-time acceleration and early inflation. There exists a limited number of scholarly works that successfully integrate the early cosmos with the contemporary universe, as shown by Scherrer (2004)\cite{Scherrer1}. In our study, an important finding has been made, indicating that our model has the capability to generate both the early universe and the current universe by considering the equation of state (EoS) parameter, denoted as $\omega$, which exhibits a transition from positive to negative values. A positive value of the parameter $\omega$ is indicative of the early universe, whereas a negative value of $\omega$ corresponds to the current or accelerated world, which is dominated by dark energy. So, it may be inferred that our investigations include a comprehensive span of the universe's history, contingent upon the varying values of the EoS parameters throughout different epochs.

In the article, we have not addressed the behavior of the additional cosmological parameters, like the Hubble parameter, the deceleration parameter, the redshift function, and so on, which can be taken into account in the near future with the observational data set. For the time being, based on all the features and attributes of the present investigation, we may conclude that our model provides possibilities for a new viewpoint of the cosmological and gravitational scenario.
\\

\section*{Acknowledgement}
A.P. and G.M. acknowledge the DSTB, Government of West Bengal, India for financial support through Grant Nos. 856(Sanc.)/STBT-11012(26)/6/2021-ST SEC dated 3rd November 2023. SR and CR express thanks to IUCAA, Pune, India for hospitality and support during an academic visit where a part of this work is accomplished. SR also gratefully acknowledges the facilities under ICARD, Pune at CCASS, GLA University, Mathura. The authors are thankful to Dr. J. L. Rosa for various helpful discussions. \\

{\bf Conﬂict of Interest:}
The authors declare no conﬂict of interest.\\

{\bf Data Availability Statement:} No Data associated in the manuscript.

\appendix

\section{Expressions of full forms of the  EoS parameters for various models of \texorpdfstring{$f(\R)$}{TEXT}}
\subsection{\texorpdfstring{$f(\R)=\R$}{TEXT}}
The expression of $\tilde{\rho}_1$ and $\tilde{p}_{1}$ is given by:
\begin{eqnarray}
\tilde{p}_{1}&&=\frac{1}{\Big(48 \lambda ^2+16 \lambda +1\Big) \dot{\phi}^2 \Big(\dot{\phi}^2-1\Big)^3}\Big[2 \Big(8 \lambda ^2 \sqrt{1-\dot{\phi}^2} \dot{\phi}^{12}-24 \lambda ^2 \sqrt{1-\dot{\phi}^2} \dot{\phi}^{10}\nonumber\\&&+2 \lambda  \dot{\phi}^8 \Big(\lambda  \Big(7 \sqrt{1-\dot{\phi}^2}-4\Big)+\sqrt{1-\dot{\phi}^2}-1\Big)+2 \lambda  \dot{\phi}^6 \Big(\lambda  \Big(12-7 \sqrt{1-\dot{\phi}^2}\Big)-3 \sqrt{1-\dot{\phi}^2}+3\Big)\nonumber\\
&&+6 \lambda  (4 \lambda +1) \Big(\sqrt{1-\dot{\phi}^2}-1\Big) \dot{\phi}^4-2 (6 \lambda +1) \dot{\ddot{\phi}} \dot{\phi}^3+2 (6 \lambda +1) \dot{\ddot{\phi}} \dot{\phi}\nonumber\\
&&+\dot{\phi}^2 \Big(8 (6 \lambda +1) \ddot{\phi}^2-2 \lambda  (4 \lambda +1) \Big(\sqrt{1-\dot{\phi}^2}-1\Big)\Big)+\ddot{\phi}^2\Big)\Big]
\label{A1}
\een
\ben
\tilde{\rho_1}&&=\frac{1}{\Big(48 \lambda ^2+16 \lambda +1\Big) \dot{\phi}^2 \Big(\dot{\phi}^2-1\Big)^3}\Big[2 \Big(-3 (8 \lambda +1) \ddot{\phi}^2+4 \lambda  (10 \lambda +1) \sqrt{1-\dot{\phi}^2} \dot{\phi}^{12}-12 \lambda  (10 \lambda +1) \sqrt{1-\dot{\phi}^2} \dot{\phi}^{10}\nonumber\\
&&+\lambda  \dot{\phi}^8 \Big(\lambda  \Big(22 \sqrt{1-\dot{\phi}^2}+8\Big)+\sqrt{1-\dot{\phi}^2}+2\Big)+\lambda  \dot{\phi}^6 \Big(\lambda  \Big(74 \sqrt{1-\dot{\phi}^2}-24\Big)+11 \sqrt{1-\dot{\phi}^2}-6\Big)\nonumber\\
&&-6 \lambda  (4 \lambda +1) \Big(\sqrt{1-\dot{\phi}^2}-1\Big) \dot{\phi}^4-12 \lambda  \dot{\ddot{\phi}} \dot{\phi}^3+12 \lambda  \dot{\ddot{\phi}} \dot{\phi}+2 \lambda  \dot{\phi}^2 \Big((4 \lambda +1) \Big(\sqrt{1-\dot{\phi}^2}-1\Big)+24 \ddot{\phi}^2\Big)\Big)\Big].
\label{A2}
\end{eqnarray}

Using Eqs. (\ref{A1}), (\ref{A2}) and (\ref{39}), we have the numerator and denominator of Eq. (\ref{40}) as:

\ben
\o_{N_1}&=&-96 \lambda ^2 \tau ^2 e^{t/\tau } \sqrt{1-e^{-\frac{t}{\tau }}}+32 \lambda ^2 \tau ^2 \sqrt{1-e^{-\frac{t}{\tau }}}-8 \lambda  \tau ^2 e^{\frac{3 t}{\tau }} \Big(-12 \lambda +7 \lambda  \sqrt{1-e^{-\frac{t}{\tau }}}+3 \sqrt{1-e^{-\frac{t}{\tau }}}-3\Big)\nonumber\\
&&+8 \lambda  \tau ^2 e^{\frac{2 t}{\tau }} \Big(\lambda  \Big(7 \sqrt{1-e^{-\frac{t}{\tau }}}-4\Big)+\sqrt{1-e^{-\frac{t}{\tau }}}-1\Big)-(4 \lambda +1) e^{\frac{5 t}{\tau }} \Big(8 \lambda  \tau ^2 \Big(\sqrt{1-e^{-\frac{t}{\tau }}}-1\Big)-3\Big)\nonumber\\
&&+6 e^{\frac{4 t}{\tau }} \Big(6 \lambda +4 \lambda  (4 \lambda +1) \tau ^2 \Big(\sqrt{1-e^{-\frac{t}{\tau }}}-1\Big)+1\Big).
\label{A3}
\een
\ben
\o_{D_1}&=&-48 \lambda  (10 \lambda +1) \tau ^2 e^{t/\tau } \sqrt{1-e^{-\frac{t}{\tau }}}+4 \lambda  \tau ^2 e^{\frac{2 t}{\tau }} \Big(\lambda  \Big(22 \sqrt{1-e^{-\frac{t}{\tau }}}+8\Big)+\sqrt{1-e^{-\frac{t}{\tau }}}+2\Big)\nonumber\\&&+4 \lambda  \tau ^2 e^{\frac{3 t}{\tau }} \Big(\lambda  \Big(74 \sqrt{1-e^{-\frac{t}{\tau }}}-24\Big)+11 \sqrt{1-e^{-\frac{t}{\tau }}}-6\Big)+16 \lambda  (10 \lambda +1) \tau ^2 \sqrt{1-e^{-\frac{t}{\tau }}}\nonumber\\
&&+(4 \lambda +1) e^{\frac{5 t}{\tau }} \Big(8 \lambda  \tau ^2 \Big(\sqrt{1-e^{-\frac{t}{\tau }}}-1\Big)-3\Big)\nonumber\\
&&-12 \lambda  e^{\frac{4 t}{\tau }} \Big(2 (4 \lambda +1) \tau ^2 \Big(\sqrt{1-e^{-\frac{t}{\tau }}}-1\Big)-3\Big).
\label{A4}
\een
\subsection{\texorpdfstring{$f(\R)=\R+\alpha\R^{2}$}{TEXT}}
The expression of $\tilde{\rho}_{2}$ and $\tilde{p}_{2}$ is given by:
\ben
\tilde{p}_{2}&=&\frac{2}{\Big(48 \lambda ^2+16 \lambda +1\Big) \dot{\phi}^4 \Big(\dot{\phi}^2-1\Big)^6}\Big[18 \alpha  (12 \lambda +1) \ddot{\phi}^4+4 (6 \lambda +1) \Big(2 \dot{\ddot{\phi}}-3 \alpha  \dot{\ddot{\ddot{\phi}}}\Big) \dot{\phi} ^9-72 \alpha  (7 \lambda +1) \dot{\ddot{\phi}} \dot{\phi} \ddot{\phi}^2\nonumber\\
&&-12 (6 \lambda +1) \dot{\phi}^7 \Big(\dot{\ddot{\phi}} \Big(124 \alpha  \ddot{\phi}^2+1\Big)-3 \alpha  \dot{\ddot{\ddot{\phi}}}\Big)+4 \dot{\phi}^5 \Big(2 \dot{\ddot{\phi}} \Big(6 \lambda +3 \alpha  (219 \lambda +32) \ddot{\phi}^2+1\Big)-9 \alpha  (6 \lambda +1) \dot{\ddot{\ddot{\phi}}}\Big)\nonumber\\
&&+2 \dot{\phi}^3 \Big(6 \alpha  (6 \lambda +1) \dot{\ddot{\ddot{\phi}}}+\dot{\ddot{\phi}} \Big(-6 \lambda +36 \alpha  (58 \lambda +11) \ddot{\phi}^2-1\Big)\Big)+\dot{\phi}^8 \Big(156 (6 \alpha  \lambda +\alpha ) \dot{\ddot{\phi}}^2+192 \alpha  (6 \lambda +1) \ddot{\ddot{\phi}} \ddot{\phi}\nonumber\\
&&-(144 \lambda +23) \ddot{\phi}^2+10 \lambda  \Big(-12 \lambda +11 \lambda  \sqrt{1-\dot{\phi}^2}+3 \sqrt{1-\dot{\phi} ^2}-3\Big)\Big)-3 \dot{\phi}^6 \Big(2 \alpha  (288 \lambda +49) \dot{\ddot{\phi}}(t)^2\nonumber\\
&&-480 (6 \alpha  \lambda +\alpha ) \ddot{\phi}^4+24 \alpha  (31 \lambda +5)\ddot{\ddot{ \phi}} \ddot{\phi}-(48 \lambda +7) \ddot{\phi}^2+4 \lambda  (4 \lambda +1) \Big(\sqrt{1-\dot{\phi}^2}-1\Big)\Big)+\dot{\phi}^4 \Big(24 \alpha  (27 \lambda +5) \dot{\ddot{\phi}}^2\nonumber\\
&&+96 \alpha  (90 \lambda +17) \ddot{\phi}^4+144 \alpha  (7 \lambda +1)\ddot{\ddot{ \phi}} \ddot{\phi}-(48 \lambda +5) \ddot{\phi}^2+2 \lambda  (4 \lambda +1) \Big(\sqrt{1-\dot{\phi}^2}-1\Big)\Big)+\dot{\phi}^2 \Big(18 \alpha  (8 \lambda +1) \dot{\ddot{\phi}}^2\nonumber\\
&&-24 \alpha  (54 \lambda +5) \ddot{\phi}^4+24 \alpha  (3 \lambda +1) \ddot{\ddot{\phi}} \ddot{\phi} -\ddot{\phi}^2\Big)\nonumber\\
&&+8 \lambda ^2 \sqrt{1-\dot{\phi}^2} \dot{\phi}^{20}-48 \lambda ^2 \sqrt{1-\dot{\phi}^2} \dot{\phi}^{18}+2 \lambda  \dot{\phi}^{16} \Big(\lambda  \Big(55 \sqrt{1-\dot{\phi}^2}-4\Big)+\sqrt{1-\dot{\phi}^2}-1\Big)-4 \lambda  \dot{\phi}^{14} \Big(-12 \lambda +\nonumber\\
&&34 \lambda  \sqrt{1-\dot{\phi}^2}+3 \sqrt{1-\dot{\phi}^2}-3\Big)+6 \lambda  \dot{\phi}^{12} \Big(-20 \lambda +22 \lambda  \sqrt{1-\dot{\phi}^2}+5 \sqrt{1-\dot{\phi}^2}-5\Big)-2 (6 \lambda +1) \dot{\ddot{\phi}} \dot{\phi}^{11}\nonumber\\
&&-8 \dot{\phi}^{10} \Big(\lambda  \Big(-20 \lambda +17 \lambda  \sqrt{1-\dot{\phi}^2}+5 \sqrt{1-\dot{\phi}^2}-5\Big)-(6 \lambda +1) \ddot{\phi}^2\Big)\Big],
\label{A5}
\een
\ben
\tilde{\rho}_{2}&=&\frac{2}{\Big(48 \lambda ^2+16 \lambda +1\Big) \dot{\phi}^4 \Big(\dot{\phi} ^2-1\Big)^6}\Big[54 \alpha  (12 \lambda +1) \ddot{\phi}^4+24 \lambda  \Big(2 \dot{\ddot{\phi}}-3 \alpha  \dot{\ddot{\ddot{\phi}}}\Big) \dot{\phi} ^9-36 \alpha  (22 \lambda +1) \dot{\ddot{\phi}}  \dot{\phi} \ddot{\phi}^2\nonumber\\
&&-72 \lambda  \dot{\phi}^7 \Big(\dot{\ddot{\phi}} \Big(124 \alpha  \ddot{\phi}^2+1\Big)-3 \alpha\dot{\ddot{\ddot{\phi}}}\Big)+12 \dot{\phi}^5 \Big(\dot{\ddot{\phi}} \Big(4 \lambda +3 \alpha  (218 \lambda +9) \ddot{\phi}^2\Big)-18 \alpha  \lambda  \dot{\ddot{\ddot{\phi}}}\Big)\nonumber\\
&&+12 \dot{\phi} ^3 \Big(6 \alpha  \lambda  \phi ^{(5)}(t)+\dot{\ddot{\phi}} \Big(12 \alpha  (13 \lambda -2) \ddot{\phi}^2-\lambda \Big)\Big)\nonumber\\
&&+\dot{\phi} ^8 \Big(936 \alpha  \lambda  \dot{\ddot{\phi}} ^2+1152 \alpha  \lambda  \ddot{\ddot{\phi}} \ddot{\phi}-3 (56 \lambda +1) \ddot{\phi} ^2-5 \lambda  \Big(\lambda  \Big(34 \sqrt{1-\dot{\phi}^2}-24\Big)+7 \sqrt{1-\dot{\phi} ^2}-6\Big)\Big)\nonumber\\
&&+3 \dot{\phi}^6 \Big(6 \alpha  (1-88 \lambda ) \dot{\ddot{\phi}} ^2+2880 \alpha  \lambda  \ddot{\phi} ^4-12 \alpha  (70 \lambda +1) \ddot{\ddot{\phi}} \ddot{\phi}+(72 \lambda +3) \ddot{\phi}^2+4 \lambda  (4 \lambda +1) \Big(\sqrt{1-\dot{\phi} ^2}-1\Big)\Big)\nonumber\\
&&+\dot{\phi}^4 \Big(36 \alpha  (10 \lambda -1) \dot{\ddot{\phi}} ^2+576 \alpha  (7 \lambda -1) \ddot{\phi}^4+72 \alpha  (22 \lambda +1) \ddot{\ddot{\phi}} \ddot{\phi}-3 (40 \lambda +3) \ddot{\phi}^2-2 \lambda  (4 \lambda +1) \Big(\sqrt{1-\dot{\phi}^2}-1\Big)\Big)\nonumber\\
&&-3 \dot{\phi}^2 \Big(-6 \alpha  (16 \lambda +1) \dot{\ddot{\phi}} ^2+48 \alpha  (25 \lambda +2) \ddot{\phi} ^4+12 \alpha  (6 \lambda +1) \ddot{\ddot{\phi}} \ddot{\phi}-(8 \lambda +1) \ddot{\phi}^2\Big)\nonumber\\
&&+4 \lambda  (10 \lambda +1) \sqrt{1-\dot{\phi} ^2} \dot{\phi}^{20}-24 \lambda  (10 \lambda +1) \sqrt{1-\dot{\phi}^2} \dot{\phi} ^{18}\nonumber\\
&&+\lambda  \dot{\phi}^{16} \Big(\lambda  \Big(502 \sqrt{1-\dot{\phi} ^2}+8\Big)+49 \sqrt{1-\dot{\phi} ^2}+2\Big)-4 \lambda  \dot{\phi}^{14} \Big(2 \lambda  \Big(49 \sqrt{1-\dot{\phi} ^2}+6\Big)+8 \sqrt{1-\dot{\phi} ^2}+3\Big)\nonumber\\
&&-6 \lambda  \dot{\phi}^{12} \Big(10 \lambda  \Big(\sqrt{1-\dot{\phi} ^2}-2\Big)+4 \sqrt{1-\dot{\phi}^2}-5\Big)-12 \lambda  \dot{\ddot{\phi}} \dot{\phi}^{11}+4 \lambda  \dot{\phi}^{10} \Big(10 \lambda  \Big(7 \sqrt{1-\dot{\phi}^2}-4\Big)+12 \ddot{\phi}^2\nonumber\\
&&+13 \sqrt{1-\dot{\phi}^2}-10\Big)\Big)\Big].
\label{A6}
\een
Using Eqs. (\ref{A5}), (\ref{A6}) and (\ref{39}), different parts of Eq. (\ref{42}) has been expressed as:
\ben
\o_{N_2}&=&6 e^{\frac{7 t}{\tau }} \Big(-15 \alpha  (24 \lambda +5)-\tau ^2+16 \lambda  (4 \lambda +1) \tau ^4 \Big(\sqrt{1-e^{-\frac{t}{\tau }}}-1\Big)\Big)-e^{\frac{6 t}{\tau }} \Big(27 \alpha  (188 \lambda +33)+18 (8 \lambda +1) \tau ^2\nonumber\\
&&+80 \lambda  \tau ^4 \Big(-12 \lambda +11 \lambda  \sqrt{1-e^{-\frac{t}{\tau }}}+3 \sqrt{1-e^{-\frac{t}{\tau }}}-3\Big)\Big)+2 e^{\frac{5 t}{\tau }} \Big(-72 \alpha  (6 \lambda +1)+3 (32 \lambda +5) \tau ^2\nonumber\\&&+32 \lambda  \tau ^4 \Big(-20 \lambda+17 \lambda  \sqrt{1-e^{-\frac{t}{\tau }}}+5 \sqrt{1-e^{-\frac{t}{\tau }}}-5\Big)\Big)+384 \lambda ^2 \tau ^4 e^{t/\tau } \sqrt{1-e^{-\frac{t}{\tau }}}-64 \lambda ^2 \tau ^4 \sqrt{1-e^{-\frac{t}{\tau }}}\nonumber\\
&&+32 \lambda  \tau ^4 e^{\frac{3 t}{\tau }} \Big(-12 \lambda +34 \lambda  \sqrt{1-e^{-\frac{t}{\tau }}}+3 \sqrt{1-e^{-\frac{t}{\tau }}}-3\Big)-16 \lambda  \tau ^4 e^{\frac{2 t}{\tau }} \Big(\lambda  \Big(55 \sqrt{1-e^{-\frac{t}{\tau }}}-4\Big)\nonumber\\
&&+\sqrt{1-e^{-\frac{t}{\tau }}}-1\Big)-2 (4 \lambda +1) \tau ^2 e^{\frac{8 t}{\tau }} \Big(8 \lambda  \tau ^2 \Big(\sqrt{1-e^{-\frac{t}{\tau }}}-1\Big)-3\Big)\nonumber\\
&&-12 \tau ^2 e^{\frac{4 t}{\tau }} \Big(6 \lambda +4 \lambda  \tau ^2 \Big(-20 \lambda +22 \lambda  \sqrt{1-e^{-\frac{t}{\tau }}}+5 \sqrt{1-e^{-\frac{t}{\tau }}}-5\Big)+1\Big),
\label{A7}
\een
\ben
\o_{D_2}&=&-6 e^{\frac{7 t}{\tau }} \Big(-45 \alpha -3 (8 \lambda +1) \tau ^2+16 \lambda  (4 \lambda +1) \tau ^4 \Big(\sqrt{1-e^{-\frac{t}{\tau }}}-1\Big)\Big)\nonumber\\
&&+e^{\frac{6 t}{\tau }} \Big(27 \alpha  (5-148 \lambda )-18 (16 \lambda +1) \tau ^2\nonumber\\
&&+40 \lambda  \tau ^4 \Big(-24 \lambda +34 \lambda  \sqrt{1-e^{-\frac{t}{\tau }}}+7 \sqrt{1-e^{-\frac{t}{\tau }}}-6\Big)\Big)\nonumber\\
&&+e^{\frac{5 t}{\tau }} \Big(6 \tau ^2+16 \lambda  \Big(-54 \alpha +15 \tau ^2+\tau ^4 \Big(20 \lambda  \Big(4-7 \sqrt{1-e^{-\frac{t}{\tau }}}\Big)\nonumber\\
&&-26 \sqrt{1-e^{-\frac{t}{\tau }}}+20\Big)\Big)\Big)+32 \lambda  \tau ^4 e^{\frac{3 t}{\tau }} \Big(2 \lambda  \Big(49 \sqrt{1-e^{-\frac{t}{\tau }}}+6\Big)+8 \sqrt{1-e^{-\frac{t}{\tau }}}+3\Big)\nonumber\\
&&-8 \lambda  \tau ^4 e^{\frac{2 t}{\tau }} \Big(\lambda  \Big(502 \sqrt{1-e^{-\frac{t}{\tau }}}+8\Big)\nonumber\\
&&+49 \sqrt{1-e^{-\frac{t}{\tau }}}+2\Big)+192 \lambda  (10 \lambda +1) \tau ^4 e^{t/\tau } \sqrt{1-e^{-\frac{t}{\tau }}}-32 \lambda  (10 \lambda +1) \tau ^4 \sqrt{1-e^{-\frac{t}{\tau }}}\nonumber\\
&&+2 (4 \lambda +1) \tau ^2 e^{\frac{8 t}{\tau }} \Big(8 \lambda  \tau ^2 \Big(\sqrt{1-e^{-\frac{t}{\tau }}}-1\Big)-3\Big)\nonumber\\
&&+24 \lambda  \tau ^2 e^{\frac{4 t}{\tau }} \Big(2 \tau ^2 \Big(10 \lambda  \Big(\sqrt{1-e^{-\frac{t}{\tau }}}-2\Big)+4 \sqrt{1-e^{-\frac{t}{\tau }}}-5\Big)-3\Big).
\label{A8}
\een
\subsection{\texorpdfstring{$f(\R,\T)=\R+\a \R^n +\l \T$}{TEXT}}
As per the similar calculations of the above types of $f(\R,\T)$ we have the EoS parameter ($\o$) for this choice of $f(\R,\T)$ can be expressed as follows $\o=\frac{\o_{N_3}}{\o_{D_3}}$ where
\ben
\o_{N_3}&&=12 (e^{-t}-1) \Bigg[2^{-n-1} 3^n \Big[\frac{e^{2 t} (2 e^t-3)}{(e^t-1)^3}\Big]^n\nonumber\\
&&+\frac{1}{(2 e^t-3)^3}\Big[2^{3-n} 3^{n-1} (n-1) n e^{-3 t} \Big\{(n-4) e^{5 t}+9 (n-2) e^t+14 (n-1) e^{3 t}+(15-6 n) e^{4 t}\nonumber\\
&&+(11-16 n) e^{2 t}-2 n+4\Big\} \Big(\frac{e^{2 t} (2 e^t-3)}{(e^t-1)^3}\Big)^n\Big]-\frac{1}{(3-2 e^t)^2}\Big\{\Big(\frac{3}{2}\Big)^n (n-1) n (e^t-2) \Big(\frac{e^{2 t} (2 e^t-3)}{(e^t-1)^3}\Big)^n\Big\}\nonumber\\
&&+\frac{1}{2 (1-e^{-t})^3}\Big[3 (e^{-t}-1) \Big\{n \Big(\frac{2}{3}\Big)^{1-n} \Big(\frac{e^{2 t} (2 e^t-3)}{(e^t-1)^3}\Big)^{n-1}+1\Big\}\Big]\nonumber\\
&&+\frac{1}{(e^t-1)^3 (2 e^t-3)}\Big[2^{-n-2} (e^t-3) \Big\{-2\ 3^n n \Big(\frac{e^{2 t} (2 e^t-3)}{(e^t-1)^3}\Big)^n+2\ 3^{n+1} n e^t \Big(\frac{e^{2 t} (2 e^t-3)}{(e^t-1)^3}\Big)^n\nonumber\\
&&+2 e^{3 t} \Big(3^n n \Big(\frac{e^{2 t} (2 e^t-3)}{(e^t-1)^3}\Big)^n+3\ 2^n\Big)-3 e^{2 t} \Big(2\ 3^n n \Big(\frac{e^{2 t} (2 e^t-3)}{(e^t-1)^3}\Big)^n+3\ 2^n\Big)\Big\}\Big]\nonumber\\
&&+\frac{e^{-2 t} \Big(-4 e^{-2 t}+8 e^{-t}+5\Big)}{2 (1-e^{-t})^{3/2}}+\frac{3 e^{2 t} (2 e^t-3)}{4 (e^t-1)^3}-\sqrt{1-e^{-t}}+1\Bigg],
\label{A9}
\een
\ben
\o_{D_3}=&&5 \Biggr[2^{-n} 3^{n+1} e^{-t} (-1+e^t) \Big[\frac{e^{2 t} (-3+2 e^t)}{(-1+e^t)^3}\Big]^n\nonumber\\&&+\frac{1}{(3-2 e^t)^2}\Bigg[2^{1-n} 3^{n+1} e^{-t} (-2+e^t) (-1+e^t) (n-1) n \Big[\frac{e^{2 t} (-3+2 e^t)}{(-1+e^t)^3}\Big]^n\Bigg]\nonumber\\&&
+\frac{9 e^t (-3+2 e^t)}{2 (-1+e^t)^2}+\frac{18}{5} (-1+e^{-t}) \Bigg[2^{-n-1} 3^n \Big[\frac{e^{2 t} (-3+2 e^t)}{(-1+e^t)^3}\Big]\Bigg]^n\nonumber\\&&
+\frac{1}{(-3+2 e^t)^3}\Bigg[2^{3-n} 3^{n-1} e^{-3 t} (n-1) n \Big[e^{5 t} (n-4)+9 e^t (n-2)+14 e^{3 t} (n-1)-2 n+e^{4 t} (15-6 n)\nonumber\\&&
+e^{2 t} (11-16 n)+4\Big]\Big[\frac{e^{2 t} (-3+2 e^t)}{(-1+e^t)^3}\Big]^n\Bigg]-\frac{1}{(3-2 e^t)^2}\Bigg[\Big(\frac{3}{2}\Big)^n (-2+e^t) (n-1) n \Big[\frac{e^{2 t} (-3+2 e^t)}{(-1+e^t)^3}\Big]^n\Bigg]
\nonumber\\&&+\frac{e^{-2 t} (5-4 e^{-2 t}+8 e^{-t})}{2 (1-e^{-t})^{3/2}}+\frac{3 e^{2 t} (-3+2 e^t)}{4 (-1+e^t)^3}
\nonumber\\
&&+\frac{1}{2 (1-e^{-t})^3}\Bigg[3 (-1+e^{-t}) \Big[\Big(\frac{e^{2 t} (-3+2 e^t)}{(-1+e^t)^3}\Big)^{n-1} n \Big(\frac{2}{3}\Big)^{1-n}+1\Big]\Bigg]\nonumber\\&&
+\frac{1}{(-1+e^t)^3 (-3+2 e^t)}\Bigg[2^{-n-2} (-3+e^t) \Big[-2~ 3^n n \Big(\frac{e^{2 t} (-3+2 e^t)}{(-1+e^t)^3}\Big)^n+2\ 3^{n+1} e^t n \Big(\frac{e^{2 t} (-3+2 e^t)}{(-1+e^t)^3}\Big)^n\nonumber\\
&&+2 e^{3 t} \Big(3^n n \Big(\frac{e^{2 t} (-3+2 e^t)}{(-1+e^t)^3}\Big)^n+3\ 2^n\Big)-3 e^{2 t} \Big(2\ 3^n n \Big(\frac{e^{2 t} (-3+2 e^t)}{(-1+e^t)^3}\Big)^n+3\ 2^n\Big)\Big]\Big)-\sqrt{1-e^{-t}}+1\Bigg]\nonumber\\&&
+6 \Big(-1+e^{-t}\Big) \Big(\sqrt{1-e^{-t}}-1\Big)-\frac{1}{2 (-1+e^t)^2}\Bigg[9 e^t (-3+e^t) \Big(\Big(\frac{e^{2 t} (-3+2 e^t)}{(-1+e^t)^3}\Big)^{n-1} n \Big(\frac{2}{3}\Big)^{1-n}+1\Big)\Bigg]\nonumber\\
&&-\frac{3 e^{-4 t} \Big(-4+8 e^t+5 e^{2 t}\Big)}{\sqrt{1-e^{-t}}}\Biggr].
\label{A10}
\een

\end{document}